\documentclass[fleqn,usenatbib]{mnras}

\usepackage{newtxtext,newtxmath}

\usepackage[T1]{fontenc}

\DeclareRobustCommand{\VAN}[3]{#2}
\let\VANthebibliography\thebibliography
\def\thebibliography{\DeclareRobustCommand{\VAN}[3]{##3}\VANthebibliography}

\usepackage{graphicx}	% Including figure files
\usepackage{amsmath}	% Advanced maths commands
\usepackage{cancel}
\usepackage{xfrac}
\usepackage{svg}
\newcommand{\orcid}[1]{\href{https://orcid.org/#1}{\includesvg[width=10pt]{orcid}}}

\defcitealias{Popham1991}{PN91}
\defcitealias{Paczynski1991}{P91}
\defcitealias{Lechien2025}{L25}
\defcitealias{Martin2025}{M25}
\defcitealias{Xing2026}{X26}

\newcommand{\mesa}{\textsc{mesa}\,}
\newcommand{\posydon}{\textsc{posydon}\,}
\newcommand{\bpass}{\textsc{bpass}\,}

\title[Disc-mediated angular momentum transport resolves the mass-gain problem in rotating binaries]{Disc-mediated angular momentum transport resolves the mass-gain problem in rotating binaries}

\author[Ljung, Gilkis \& Tacchella]{
Sebastian Ljung$^{\orcid{0000-0002-3354-1589}}$,$^{1,2}$ \thanks{\href{Sebastian.Ljung@unige.ch}{sebastian.ljung@unige.ch}}
Avishai Gilkis$^{\orcid{0000-0001-8949-5131}}$ $^{1}$\thanks{\href{agilkis@ast.cam.ac.uk}{agilkis@ast.cam.ac.uk}} and
Sandro Tacchella$^{\orcid{0000-0002-8224-4505}}$ $^{3,4}$
\\
$^{1}$Institute of Astronomy, University of Cambridge, Madingley Road, Cambridge CB3 0HA, UK\\
$^{2}$Department of Astronomy, University of Geneva, Chemin Pegasi 51, 1290 Versoix, Switzerland\\
$^{3}$Kavli Institute for Cosmology, University of Cambridge, Madingley Road, Cambridge CB3 0HA, UK\\
$^{4}$Cavendish Laboratory, University of Cambridge, 19 JJ Thomson Avenue, Cambridge CB3 0HE, UK
}

\date{Accepted XXX. Received YYY; in original form ZZZ}

\pubyear{\the\year{}}

\begin{document}
\label{firstpage}
\pagerange{\pageref{firstpage}--\pageref{lastpage}}
\maketitle

\begin{abstract}

A large fraction of stars interact with a close companion during their lifetime, in which the transfer of mass and angular momentum shapes their evolution and final fate. Standard rotationally limited accretion models predict that accretors reach critical rotation after gaining only a small fraction of their mass, severely suppressing further accretion. This is in tension with observations of post-interaction systems that require substantial mass gain. We introduce a disc-mediated angular momentum transport prescription for mass-transfer onto stellar companions. This is based on a novel perturbative, analytic star--disc boundary model that allows for continued mass inflow as the accretor approaches critical rotation. Furthermore, the model reproduces previous numerical results in which perturbations to super-critical rotation result in negative torques exerted by the disc, extracting excess angular momentum while allowing continued mass inflow. We implement this mechanism in detailed binary evolution calculations with differential rotation using \textsc{mesa}, and compute grids spanning primary mass, orbital period, and mass ratio. Whereas rotationally limited models predict $\beta_{\rm eff}\lesssim0.1$, the disc model yields sustained mass inflow near critical rotation, with effective mass-transfer efficiencies of $\beta_{\rm eff}\sim0.4$–$1$ across much of the parameter space. The resulting accretor properties are broadly consistent with observed post-interaction sdOB+Be binaries. Disc-mediated angular momentum transport may therefore represent a key missing ingredient in standard binary evolution models, with important implications for rapidly rotating stars and compact-object progenitors.

\end{abstract}


%%%%%%%%%%%%%%%%%%%%%%%%%%%%%%%%%%%%%%%%%%%%%%%%%%

\section{Introduction} \label{sec:ITS}

Stellar binary interactions, such as Roche-lobe overflow (RLOF) mass-transfer or coalescence, play a central role in shaping the evolution, structure, and final fate of stars. A large fraction of all stars are born in binary systems \citep{Duchene2013,Moe2017}, where for massive stars, interaction with a companion during their lifetime is more likely than not (\citealt*{Vanbeveren1998}; \citealt{Sana2012,Sana2025}). These interactions influence their lifetimes, nucleosynthetic yields, spectra, and endpoints as compact objects or supernova (SN) progenitors (e.g. \citealt*{Podsiadlowski1992}; \citealt{deMink2013,deMink2014,Gilkis2019}). Accurately modelling mass and angular momentum transfer in interacting binaries is therefore essential for interpreting stellar populations, transient phenomena, and gravitational-wave sources.

Modern one-dimensional stellar evolution codes include the treatment of differential rotation, internal angular momentum transport, and rotationally induced mixing (e.g. \citealt{Eggenberger2008}; \citealt*{Potter2012}; \citealt{Paxton2013}). This enables the modelling of binary interactions (including tides) simultaneously with differential rotation (e.g. \citealt{Paxton2015}), however, modelling mass-transfer onto rapidly rotating accretors remains a major challenge. In practice, numerical difficulties associated with the accumulation of angular momentum at the stellar surface have led many binary calculations either to neglect rotation altogether or to adopt simplified prescriptions that severely limit mass gain once the accretor approaches critical rotation.

When considering mass gain through RLOF, two key physical aspects should be taken into account: thermal changes in the accretor and its spin-up towards critical rotation, both potentially resulting in non-conservative mass-transfer such that not all of the mass lost from the donor, is gained by the accretor. Two classes of prescriptions are therefore commonly employed for estimating the mass-transfer efficiency. Thermally limited accretion assumes that accretion faster than the Kelvin--Helmholtz timescale, $\tau_\mathrm{KH}$, drives the accretor out of thermal equilibrium, rendering the accreted material loosely bound and susceptible to removal via winds. This is typically implemented by capping the accretion rate at $\dot m_\mathrm{therm}= M / \tau_\mathrm{KH}$ or some multiple thereof. Rotationally limited accretion, motivated by the rapid spin-up of the accretor \citep{Packet1981}, assumes that once the stellar surface reaches critical rotation, further accretion is suppressed, often implicitly through strongly enhanced mass loss. In practice, this leads to low mass-transfer efficiencies for wide initial orbits.

To reduce the uncertainty on mass-transfer efficiency, comparisons between observations and theoretical models are valuable. For example, blue straggler stars are hypothesised to appear younger than they actually are due to mass gain \citep{Geller2011,Mathieu2025}. Recent analyses of post-interaction systems, including sdOB+Be binaries, indicate that accretors can gain several solar masses while remaining at subcritical or near-critical rotation, thus implying higher efficiencies than predicted by rotationally limited models \citep[][hereafter L25]{Lechien2025}. Similar tensions arise in hydrogen-deficient systems such as $\upsilon$~Sagittarii, where the secondary appears to have more than doubled its mass through accretion despite rapid rotation \citep*{Gilkis2023,Bour2025}. At the population level, binary evolution calculations may also overpredict the occurrence of contact and common-envelope events, particularly at low mass ratios, suggesting that some stabilising mechanism may be missing from standard treatments \citep*{Henneco2024}.

A key uncertainty underlying these issues is the treatment of angular momentum removal from the accretor. Crucially, although the recently accreted material is limited to critical rotation, as the accretor's moment of inertia, radiative pressure, and surface rotation evolves due to mass gain, its surface may be perturbed from critical to super-critical rotation, thereby necessitating a means to remove excess angular momentum. In compact objects such as white dwarfs and neutron stars, strong magnetic fields may truncate the inner disc whilst also providing an efficient channel for angular momentum extraction. For main-sequence (MS) stars, however, such strong fields are generally not observed, and magnetic star--disc coupling could prove insufficient to regulate spin-up in many systems \citep{Bour2025, Mimes2016}. As a result, rotationally boosted winds have often been invoked as an alternative mechanism, despite significant theoretical and observational uncertainty regarding their efficiency, geometry, and angular momentum loss rates \citep*{Heger2000,Muller2014,Hastings2023}.

An alternative possibility is that angular momentum can be redistributed through the accretion disc itself. \citet[][hereafter P91]{Paczynski1991} demonstrated that when the stellar surface approaches critical rotation, the sharp boundary between the star and the disc transitions into a continuous star--disc system. In this regime, and in the absence of inner disc truncation by magnetic fields, viscous transport across the boundary becomes possible, and the net angular momentum flux can become negative if the stellar surface rotates faster than the inner disc. This inspired the model presented in this paper, and a similar model by \citet[][hereafter X26]{Xing2026}, as potential mechanisms for continued mass accretion near critical rotation. Similar physics is thought to operate in Be-star decretion discs, where angular momentum is transported outward without requiring strong winds or magnetic fields \citep[][hereafter M25]{Martin2025}.

In this paper, we develop and implement an analytic prescription inspired by the Paczy\'nski star--disc numerical model, allowing for disc-mediated angular momentum extraction when the accretor approaches critical rotation. This prescription enables continued mass accretion while transferring excess angular momentum away from the stellar surface, potentially back into the binary orbit. By incorporating this mechanism into the \textsc{mesa} code, we enable stable, continuous modelling of differentially rotating binaries from the zero-age main-sequence (ZAMS) through to binary compact-object formation.

The structure of the paper is as follows. In Section\,\ref{sec:ADM} we derive the analytic disc prescription and in Section\,\ref{sec:NM} describe its numerical implementation. In Section\,\ref{sec:PAD} we apply the prescription to a representative binary system and compare the resulting mass-transfer efficiencies to commonly used accretion models and to observational constraints. In Section\,\ref{sec:discussdiscs} we discuss the broader implications of disc-mediated angular momentum transport for binary evolution and outline directions for future work and in Section\,\ref{sec:conc} we conclude.

\begin{figure*}%
    \vspace{-3mm}

    \begin{minipage}{1.0\textwidth}

	\includegraphics[width=1\linewidth,trim={0 4cm 0 4.95cm},clip]{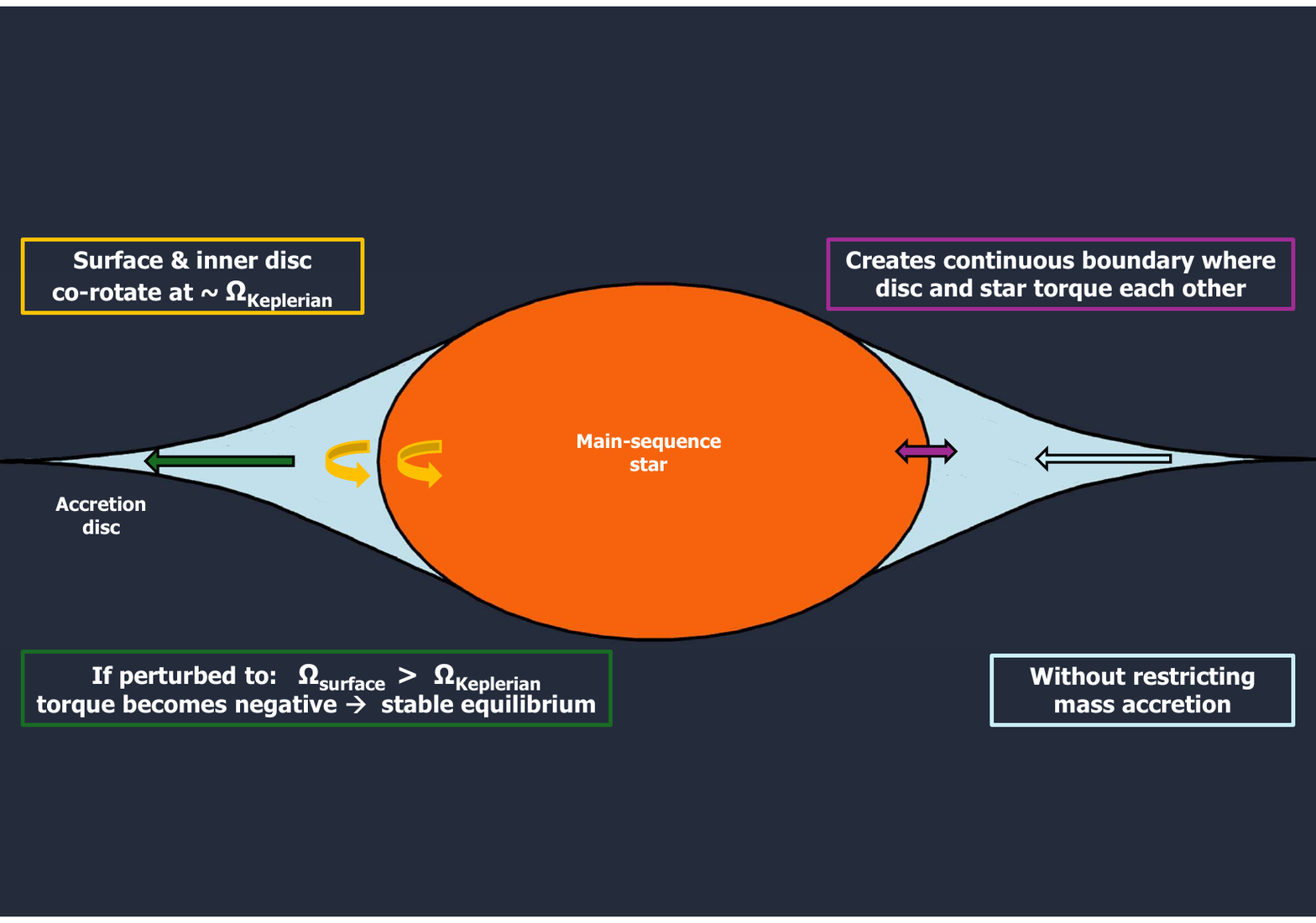}

    \end{minipage}
    \vspace{-1mm}
    \caption{Illustration of co-rotation between the inner edge of a disc and the surface of a rapidly rotating star, resulting in a continuous transition between the star and disc. The disc height is not illustrated to scale.}
    \label{fig:star-disc_illustration}
\end{figure*}

\section{Accretion disc model}
\label{sec:ADM}

\subsection{Outline of disc derivation}

We derive an accretion disc model that will allow us to follow the angular momentum evolution in accreting main-sequence companions in binary systems, as depicted in Fig.\,\ref{fig:star-disc_illustration}. In particular, we are interested in replicating the numerical models derived by \citetalias{Paczynski1991} and \citet[][hereafter PN91]{Popham1991}, who report negative torque arising when the star reaches super-critical rotation, despite positive mass inflow, $\dot m$. Such behaviour might provide an alternative source of torque, to stellar winds, in non-magnetised stars, thereby facilitating continued mass accretion at critical rotation without removing excessive amounts of mass. We begin by solving for the angular velocity $\Omega$ using the equation for conservation of angular momentum,
\begin{equation}
\frac{\partial \Omega}{\partial r} = -\frac{\dot m \Omega}{2\pi \nu \Sigma r} + \frac{\dot{J}}{2\pi \nu \Sigma r^3}\, ,
\label{eq:conservj}
\end{equation} 
where \(\nu\) is the kinematic viscosity, \(\Sigma\) is the surface density, and $\dot{J}$ is the angular-momentum flux through the inner disc. For $\dot{J}_\mathrm{Kep}=\dot m\sqrt{GMR_\star}$, where $R_\star$ is the stellar radius, a general solution for a Keplerian disc around a star is obtained. However, such models do not account for the rotation of the star, inspiring the introduction of numerical polytropic accretion disc models by \citetalias{Paczynski1991} and \citetalias{Popham1991}, which instead find that for stars near critical rotation, a second set of solutions arises that remove the no-torque boundary condition and thereby facilitate a continuous transition from the disc angular velocity $\Omega_\mathrm{Kep}\equiv \sqrt{\frac{GM}{r^3}}$ to the stellar surface angular velocity $\Omega_\star $.

A full derivation is given in Appendix\,\ref{section:appendixEquationDerivation}, and here we summarise the key results. Rearranging Eq.\,\ref{eq:conservj},
\begin{equation*}
\nu \Sigma\frac{\partial \Omega}{\partial r} = -\frac{\dot m \Omega}{2\pi  r} + \frac{\dot J}{2\pi r^3}\, ,
\end{equation*}
we isolate $r$ and $\Omega$ on the right-hand side. As we are primarily interested in the disc behaviour as it approaches critical rotation or is perturbed to super-critical rotation, we can isolate an approximate analytic solution in the regime where $\Omega^2(r)-\Omega_\mathrm{Kep}^2(r)$ is small for all $r$. This significantly simplifies the radial momentum equation, 
\begin{equation}
u_r \frac{\partial u_r}{\partial r} = - \frac{1}{\rho}\frac{\partial P}{\partial r} + r \bcancel{\left(\Omega^2 - \Omega_\mathrm{Kep}^2\right)}\, ,
\label{eq:radmom}
\end{equation}
by neglecting the small difference between centrifugal and gravitational terms\footnote{In Appendix\,\ref{section:appendixEquationDerivation}, we check this by taking numerical values from an accretor in one of our simulations undergoing mass-transfer. For smaller values, such as $|1-\left(\Omega / \Omega_\mathrm{Kep}\right)^2| \leq 0.13$, the angular velocity terms on the right-hand side contribute less than $1\,\%$ of the two terms on the left-hand side of Eq.\,\ref{eq:radmom}, thus making it effectively negligible whilst allowing for a range of $0.93 \leq \frac{\Omega}{\Omega_\mathrm{Kep}} \leq 1.066$ in which this approximation holds. This is a small but sufficient regime for our analytic model to torque the star, as demonstrated in \citetalias{Paczynski1991}'s examples. However, as this represents a singular instance, further testing is needed.}.

Using this approximation we identify a unique solution and find that its polytropic index makes $\nu \Sigma$ independent of r, allowing us to invert the angular momentum equation to isolate the following solution:
\begin{equation*}
\Omega = \Omega_\mathrm{Kep} - \frac{3\dot J}{\dot m r^2}\, .
\end{equation*}
This relation analytically ties deviations from $\Omega_\mathrm{Kep}$ at the stellar surface to $\dot J$,
\begin{equation}\label{eq:1}
    \Omega_\mathrm{Kep} - \Omega_\star = \frac{3\dot J}{\dot m R_\star^2}\, ,
\end{equation}
a result that is discussed and found in the numerical models by \citetalias{Popham1991} and \citetalias{Paczynski1991}.

We recover the behaviour that for $\Omega_\star > \Omega_\mathrm{Kep}$, if $\dot m$ is known to be positive, $\dot J$ becomes negative. Continued mass accretion is facilitated whilst applying a negative torque on the star, thus providing an alternative mechanism to magnetic disc coupling or stellar winds for removing excess angular momentum in super-critically rotating stars. As a consequence, we can conveniently model the disc torque using only one equation, making it computationally inexpensive to integrate it into our stellar evolution calculations. The model does not account for various physical aspects of real stellar accretion discs, including stellar radiative pressure on the disc, outer disc truncation by the donor star, and uncertainties regarding the thin disc approximation (or slim disc, as we include the $\frac{\partial p}{\partial r}$ term). However, many of the assumptions made are not uncommon in the field, with the exception of $\Omega_\mathrm{Kep}^2 -\Omega_\star^2$ being small but not zero. As we limit our application to near critically rotating stars perturbed into super-critical rotation, we remain within the regime in which this approximation is valid.

The derivation results in a specific polytropic index and a set of radial dependencies for each variable that can be compared to those of other published stellar disc models. We get a polytropic index of $n=\frac{3}{2}$, which aligns with the index assumed by \citet{Paczynski1991} but disagrees with that assumed in \citet{Popham1991}, $n=1$. Interestingly, we get a disc height scaling of $z_h \propto r$, which not only agrees with the numerical dependency in \citetalias{Paczynski1991}, but also with the defined constant $\frac{H}{R}$ in \citetalias{Martin2025}. Notably, this height scaling relation is not reproduced for other polytropic indices, as seen in \citetalias{Popham1991} and the isothermal model of \citet*{Dong2021}. 

As our polytropic index analytically arises within the perturbative regime of small $\Omega^2(r)-\Omega_\mathrm{Kep}^2(r)$, our model behaves smoothly near the transition between positive and negative $\dot J$. This presents advantages compared to applying the same mechanism for different polytropic indices, like that in \citet{Popham1991}, as we avoid sharp transitions self-consistently without introducing unconstrained parameters.

We also find that the speed of sound scales as $c_s \propto r^{-\frac{1}{2}}$, which aligns with that of \citetalias{Martin2025},
\begin{equation*}
    c_s \propto \frac{H}{r} r\, \Omega_\mathrm{Kep} \propto r \sqrt{\frac{GM}{r^3}} \propto r^{-\frac{1}{2}}\, .
\end{equation*}
and by relation to $c_s$ we also align on the $\nu$ dependence on $r$, but differ on $\Sigma$, for which they find a more complex relation that accounts for the disc transition radius.

The variables most important to our disc model derivation are the kinematic viscosity, $\nu \propto r^{\frac{1}{2}}$ and disc surface density, $\Sigma \propto r^{-\frac{1}{2}}$. Combined, these result in $\nu\Sigma$ being independent of $r$, which is the derived property which allows the angular-momentum equation to be solved analytically. 

\subsection{Transfer of angular momentum from the disc to the orbit} \label{subsec:MTT}

A key distinction between the disc-based model and existing implementations that set an implicit upper limit on $\Omega$ is related to whether the removed angular momentum is lost from the binary system (e.g. by stellar winds) or reinserted into the orbital angular momentum $J_\mathrm{orb}$ (e.g. through tidal interactions).

We opt for the latter approach, which we conceptually justify by assuming that the donor star applies a torque on the outer disc, which facilitates angular momentum transfer to $J_\mathrm{orb}$. This also implies that the disc would be truncated within the Roche-lobe (RL) radius of the accretor, suggesting that the mass at the outer edge of the disc would rarely gain sufficient angular momentum or energy to escape the accretor's RL. We note that our analytic model does not account for the resulting outer boundary condition self-consistently. Thus, in the current model, not accounting for radiative ablation or disc overflow through the $\mathrm{L}_2$ point, the angular momentum removed from the accretor stays within the binary system, with limited mass loss from the disc. This differs from the implementation of \citetalias{Xing2026} where either none or half of the disc removed angular momentum is returned to $J_\mathrm{orb}$, and the rest is removed from the system.

The transfer of angular momentum back into $J_\mathrm{orb}$, as opposed to removal from the system (e.g. by stellar winds), provides a potential counter-balance to unstable mass-transfer. This works by countering the decrease in $J_\mathrm{orb}$ during mass-transfer onto a less massive accretor star, in addition to the contraction in donor RL radius, in a way that rapidly and linearly scales with the amount of mass transferred. However, as the accretor's surface must first be sufficiently spun up to critical rotation, in some cases the disc might activate too late to counteract the intensive initial stage of mass-transfer. This implementation may result in higher mass-transfer efficiencies in comparison to thermally limited accretion models.

\section{Numerical method}
\label{sec:NM}

We use \textsc{mesa} version 24.08.1 \citep{Paxton2011,Paxton2013,Paxton2015,Paxton2018,Paxton2019,Jermyn2023} for the modelling of our binaries and perform the data analysis using the Python \texttt{mesa\_reader} package from \citet{Wolf2017}. We adopt the Solar metallicity value obtained by \citet*{Lodders2025}, $\mathrm{Z}_{\odot}=0.0185$. We use the $\texttt{v\_flag}$, set $\texttt{Pextra\_factor} = 0$ and use the reaction network \texttt{approx21.net}. For our binaries we model both stars, deactivate wind mass-transfer and do not follow systems transferring mass onto compact remnants, nor going through a common envelope evolution stage. We use a modified version of the Kolb mass-transfer scheme \citep{Kolb1990, Marchant2021}, and allow both ballistic \citep{Ulrich1976}, and Keplerian angular momentum accretion, in addition to a new tidal prescription that applies different synchronisation timescales to different layers depending on whether they are convective or radiative. Notable non-default choices in our inlist parameters are detailed in Appendix\,\ref{section:appendixInlist}. We elaborate on specific aspects of our implementation in the following subsections.

\subsection{Implementing the disc model in \textsc{mesa}}
\label{subsec:ITD}

Our \texttt{other\_accreted\_material\_j} module implements the disc mechanism by removing angular momentum, at super-critical rotation, near the surface via \texttt{remove\_j} in a manner that distributes the shear (based on \texttt{adjust\_J\_lost}).
In the numerical implementation we replace $\Omega_\mathrm{Kep}$ with $\Omega_\mathrm{crit}$ to account for radiative effects, assuming the same functional form applies in this modified critical regime, where $\Omega_\mathrm{crit} = \Omega_\mathrm{Kep}\sqrt{1-\Gamma_\mathrm{Edd}}$, and $\Gamma_\mathrm{Edd}\equiv L / L_\mathrm{Edd} $ is the Eddington factor, i.e. the ratio of radiative to gravitational acceleration.
We adopt a conservative threshold for critical rotation, defined as $0.9\, \Omega_\mathrm{crit}$.
We also use this to adjust the maximum specific angular momentum of the accreted mass to 
$\mathrm{min}(0.9 \,j_\mathrm{crit}(R_\star),\, \mathtt{accreted\_material\_j})$. 

When the surface exceeds $0.9\, \Omega_\mathrm{crit}$, based on Eq.\, \ref{eq:1}, we apply removal of angular momentum per unit time of 
\begin{equation*}
\mathtt{ODE\_Jdot} = (0.9\,\Omega_\mathrm{crit}  - \Omega_\star) \frac{\dot m R_\star^2}{3} = -\dot{J}_{\mathrm{orb}}\, ,    
\end{equation*}
where
\begin{equation*}
\Omega_{\rm ratio} \equiv \Omega/\Omega_{\rm crit}\, .     
\end{equation*}

In the Keplerian accretion mode, the recently accreted material has an angular velocity of $\Omega_\mathrm{crit}$, for both subcritical and critically rotating stars. As we are not modelling a steady-state system, the recently accreted mass can change the moment of inertia, radius, or $\Gamma_\mathrm{Edd}$ of the star, causing the surface (recently accreted material) to exceed $\Omega_\mathrm{crit}$, despite accreting mass at $\Omega_\mathrm{crit}$.

To reproduce the equilibrium of the star--disc system at critical rotation, we need to account for the star--disc system $\dot{\mathrm{J}}$ regulating the net torque transport within the accretion disc. For slowly rotating stars, where the no-torque boundary condition applies, the disc's only angular momentum transport is advective, whereby the newly accreted material is added with a specific angular momentum, $\mathtt{accreted\_material\_j}$. As the star is spun-up, a competing (negative) viscous angular momentum starts removing angular momentum such that 
\begin{equation*}
\dot{J}_\mathrm{net} = \mathtt{ODE\_Jdot} = \dot{m} \times \mathtt{accreted\_material\_j} + \dot{J}_\mathrm{visc},
\end{equation*}
and the net angular momentum transport is defined by the disc equation. The angular momentum removed from the star via the disc becomes
\begin{equation*}
\dot{J}_\mathrm{visc} = \mathtt{ODE\_Jdot} - \dot{m} \times \mathtt{accreted\_material\_j}.
\end{equation*}
To numerically transition between the slowly rotating and near-critically rotating regimes we linearly interpolate between the no-torque solution, where $\dot{J}_\mathrm{visc}=0$, at $\Omega_\star = 0.5\,\Omega_\mathrm{crit}$, to our solution at the numerically assigned critical rotation $\Omega_\star = 0.9\,\Omega_\mathrm{crit}$. 

The removed angular momentum is deposited into $J_\mathrm{orb}$. We note that for direct ballistic accretion streams we adjust the angular velocity of the accreted material, but assume that a disc still forms if the surface reaches critical rotation, whether this happens via material ejected through a decretion disc, or settling of unaccreted material.

A star can be perturbed to super-critical rotation at some point after mass-transfer ends due to changes in moment of inertia arising from internal structure changes. Current wind based prescriptions avoid super-critical rotation by boosting stellar winds, e.g. as implemented by \citet{Heger2000} who adopt $\dot m_\mathrm{rot} = \dot m (1-\frac{\Omega}{\Omega_\mathrm{crit}})^{-0.43}$. As an alternative, we modify the presented analytic model to approximate the effect of angular momentum removal via a decretion disc to handle these numerically problematic situations. 

If the stellar surface goes significantly above critical rotation ($>1.1\, \Omega_\mathrm{crit}$) outside mass-transfer, we assume that any super-critically rotating mass is temporarily ejected equatorially into a decretion disc. This disc is then re-accreted, during which it can apply a torque on the star. This is implemented using the modified equation
\begin{equation*}
\mathtt{ODE\_Jdot} = (0.9\,\Omega_\mathrm{crit} - \Omega_\star) \frac{ R_\star^2}{3} \frac{\left(-m_\mathrm{supercritical}\right)}{\tau_\mathrm{therm}}\, ,     
\end{equation*}
where $m_\mathrm{supercritical}$ is the mass of material that is above $0.9\,\Omega_\mathrm{Kep}(r)$ at a given time step, and we assume that this occurs on a thermal timescale,
\begin{equation*}
\tau_\mathrm{therm}=\frac{GM^2}{2RL}\, .    
\end{equation*}
We note that in the current implementation we use $m_\mathrm{supercritical}$ to estimate the disc angular momentum removal, but do not remove $m_\mathrm{supercritical}$ from the star. Based on observations of Be decretion discs, we assume that this mass is eventually reaccreted via the disc, which transports angular momentum outwards with minimal mass loss, as we currently do not account for radiative ablation or disc-mediated $\mathrm{L}_2$-overflow mass loss. We note that the implementation for this is rudimentary and further investigation is needed to determine whether this disc model could also be used to model the behaviour of Be stars, what other mechanisms need to be accounted for, and what timescales and magnitudes would be needed to reproduce the observed lifespans of Be-star discs. As the mechanism only torques the star in the absence of ongoing mass-transfer, for angular velocities $>1.1\, \Omega_\mathrm{crit}$, it does not affect the stellar evolution outside this extreme regime, for which significant uncertainties remain regardless of prescription.

\subsection{Modifications to subroutines and the omega\_mix\_solver}

We use a thermal limiting accretion prescription following \citet{Gilkis2019}, where the mass-transfer efficiency parameter $\beta$ is updated every time step based on $\beta_\mathrm{therm} = \min(\frac{\dot{m}_\mathrm{therm}}{\dot{m}}, 1)$, and additionally limit mass-transfer as the accretor approaches filling its Roche lobe, using $\beta_\mathrm{RRL}$. We adjust the computation of $\beta_\mathrm{RRL}$ from \citet{Gilkis2019} to follow a smooth decrease from 0.9999 to 0 in the range $0.4 < \frac{r}{R_\mathrm{RL}} < 0.95$. The mass-transfer efficiency is then given by $\beta = \min(\beta_\mathrm{therm},\beta_\mathrm{RRL})$. The unaccreted mass, $(1-\beta)\,\dot m$, after being removed from the donor's surface, is removed from the system as an isotropic wind.

Angular momentum transport and mixing follow an extension of the Tayler-Spruit dynamo \citep{Spruit2002} prescription by \citet*{Fuller2019}, with some performance optimisations to alleviate bottlenecks.

The \texttt{solve\_omega\_mix.f90} solver was tweaked as if it runs out of substeps without reaching \texttt{dt} (using substeps of $\mathtt{dt}$, $\mathtt{dt_\mathrm{mix}}$) it would set
\begin{equation*}
\mathtt{dt_\mathrm{mix}}=\mathtt{dt_\mathrm{remaining}}\, ,    
\end{equation*}
resulting in one big step that was more likely to fail. We increased $\mathtt{max\_steps}$, set 
\begin{equation*}
\mathtt{dt_{mix}}= \frac{\mathtt{dt_\mathrm{remaining}}}{\mathrm{\mathtt{max\_steps}-\mathtt{step}}}    
\end{equation*}
once \texttt{step} reaches $70\,\%$ of $\mathtt{max\_steps}$, in addition to adjusting $\mathtt{dt_\mathrm{mix}}$ based on \texttt{tol\_correction\_max}, \texttt{tol\_correction\_norm}, \texttt{num\_iters}, and the error magnitude in the previous substep. These changes aimed to make the solver more sensitive to errors in previous substeps in addition to avoiding a single large $\mathtt{dt_\mathrm{remaining}}$ if it runs out of steps.

We noted a sharp radial expansion in the accretor when spun up (independent of the expansion driven by thermal inequilibrium), associated with the \texttt{fp\_rot} prescription in \textsc{mesa}. This would frequently cause the accretor to fill its Roche-lobe and expand more than expected for thermal bloating. We suspect that this expansion is unphysical and likely results from a breakdown in the (perturbative) \texttt{fp\_rot} approximation near critical rotation. We opt to conservatively limit \texttt{fp\_rot} within $0.8$--$1$. Our testing found that this limit prevents the noted radial expansions and numerical instabilities, whilst still accounting for some of the perturbative rotation driven effects on the internal thermodynamics and mixing. Preliminary tests indicate that only limiting \texttt{fp\_rot} near the surface, as is done in various \texttt{test\_suite} implementations, may suffice in limiting excess radial expansions.

\begin{figure*}
    \centering    
    \begin{minipage}{0.48\textwidth}
    \hspace{-6mm}
    \includegraphics[width=1.056\linewidth,trim={0.0cm 0.2cm 0.9cm 0.1cm},clip]{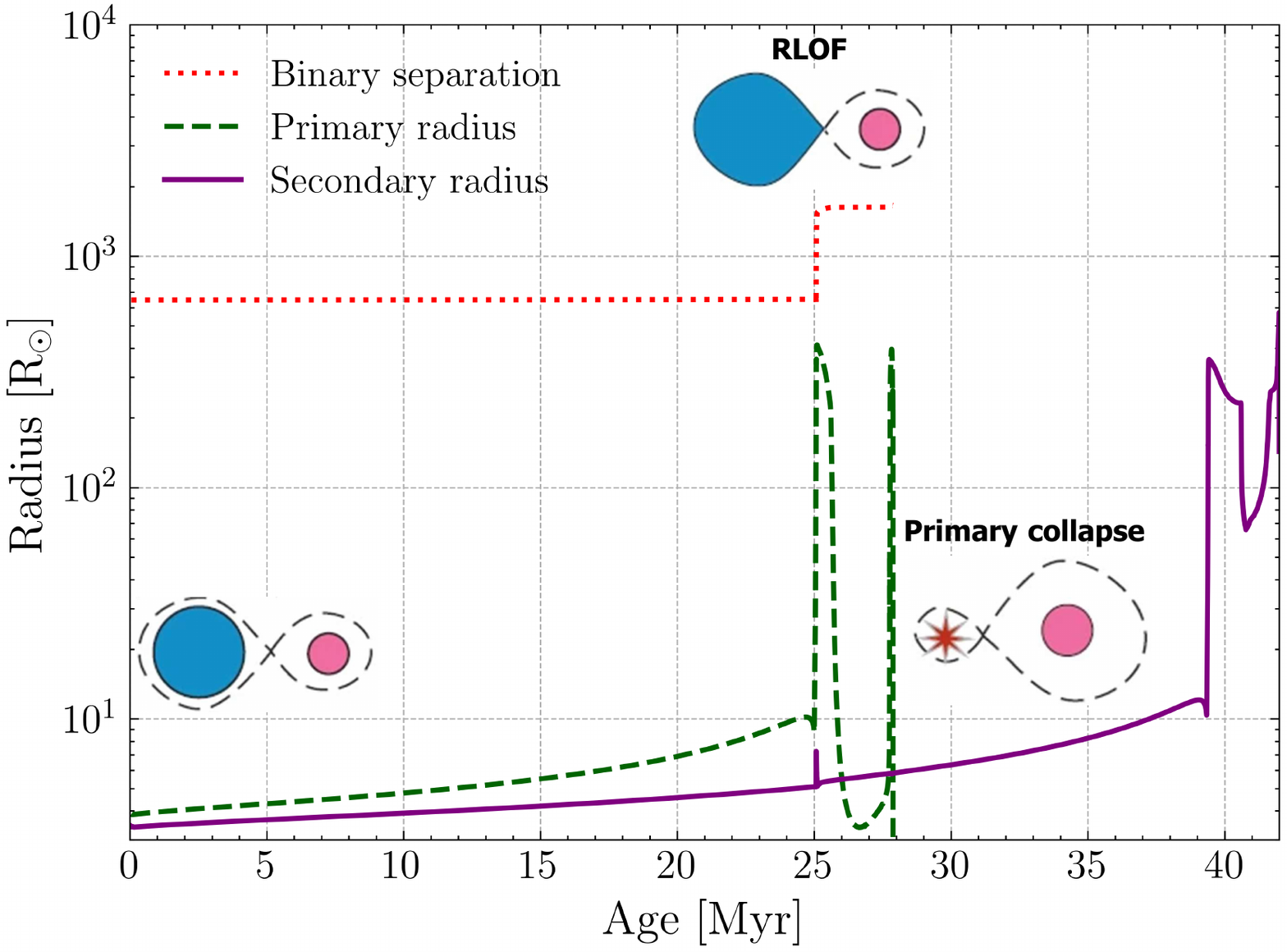}
    \end{minipage}
    \hfill
    \begin{minipage}{0.48\textwidth}
    \hspace{-3.5mm}
    \includegraphics[width=1.05\linewidth,trim={0 0.0cm 0 0.0cm},clip]{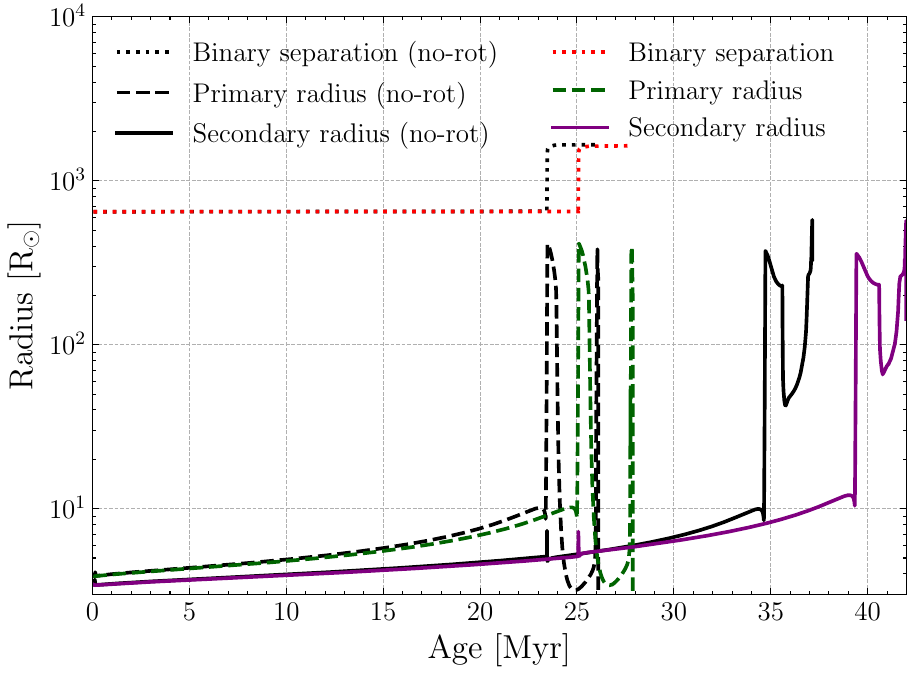}%\vspace{0.2cm}
    \end{minipage}
    \caption{\textit{Left:} The evolution of the primary radius (dashed green), secondary radius (purple) and binary separation (dotted red) with age. Roche-geometry illustrations depict the system configuration at each stage (adopted from \citealt{Postnov2006}). The primary expands to fill its RL around $25\,\mathrm{Myr}$, initiating mass transfer. The separation initially decreases until the mass ratio reverses, after which it increases. Once most of the primary's hydrogen envelope is stripped, it contracts, followed by a brief expansion before collapsing to become a compact remnant. The secondary also experiences radial expansion, but only after the binary has been disrupted by the SN explosion of the primary star. \textit{Right:} The evolution compared with that of a non-rotating equivalent (in black).}
    \label{fig:R-Age_M10P450_illustration}
\end{figure*}

\section{Numerical results and behaviour}
\label{sec:PAD}

\subsection{Detailed evolution of a binary system with the disc model} \label{DEO}

We demonstrate the effects of our disc model on stellar evolution with a detailed analysis of a $10\,\mathrm{M}_{\odot}$ + $8\,\mathrm{M}_{\odot}$ binary system with an initial orbital period of $450\,\mathrm{d}$ which will be used as a reference model in this section. In Fig.\,\ref{fig:R-Age_M10P450_illustration} (left panel) we present the evolution of the orbital separation and both stellar radii.
The primary (green-dashed line) first transfers mass onto the secondary (purple line) upon filling its Roche-lobe, and eventually collapses into a compact object while the secondary continues its main-sequence evolution\footnote{In some cases, the secondary can eventually fill its RL and transfer mass back onto the primary remnant.}.

The primary's radius increases sharply when it evolves off the MS and hydrogen burning takes place in a shell. This triggers mass-transfer during which some of the orbital angular momentum ($J_\mathrm{orb}$) is transferred to the secondary's rotation, some is removed from the binary system due to non-conservative mass-transfer (via the $\beta_\mathrm{therm}$ mechanism), and some is returned to $J_\mathrm{orb}$ via the disc mechanism. The resulting changes in separation and masses can alter the RL radius of the primary, which can cause the transfer to accelerate or decelerate, thereby altering how intensive and stable the resulting mass-transfer is.
The separation transiently decreases at the onset of mass-transfer (not discernible in Fig.\,\ref{fig:R-Age_M10P450_illustration}) until the mass ratio $q$ reaches unity, after which it increases.

After the primary has transferred a significant fraction of its mass, it eventually contracts and continues fusion of increasingly heavy elements. Meanwhile, the secondary experiences an extended main-sequence owing to the ingestion of hydrogen into its core by rotational mixing. During late stages, the primary experiences shell helium burning (or core carbon burning), potentially causing another brief expansion and mass transfer. After the primary completes core carbon burning, we assume a neutron star (NS) forms in a SN explosion, resulting in a kick which we incorporate within a \mesa timestep. For low masses the primary becomes a white dwarf rather than a NS, while for high masses a black hole can form. At this point, the separation increases rapidly (instantaneously in our modelling) and the primary mass decreases according to our SN prescription\footnote{The NS mass is determined according to a polynomial fit found by \citet{Gilkis2025}.}, after which the primary remnant is modelled as a point mass. Eventually, the secondary will also evolve off the MS, which in cases where the SN does not disrupt the binary can result in mass-transfer from the secondary back onto the remnant primary. The secondary and remnant primary are modelled until the secondary reaches white dwarf conditions, core carbon depletion, or initiates mass-transfer onto the primary remnant. Thus, we encapsulate the evolution of both primary and secondary from start to finish in one continuous run. 

A commonly-used more numerically stable alternative to including differential rotation in binary evolution is to approximate both stars as non-rotating \citep[e.g.][]{Gilkis2025}. Therefore, we compare our results to non-rotating equivalents. In the right panel of Fig.\,\ref{fig:R-Age_M10P450_illustration}, we show that the rotating and non-rotating (shown in black) model radii trace each other reasonably closely with slight deviations, with a notable exception that the rotating stars experience extended lifespans compared to their non-rotating counterparts, as a result of hydrogen ingested into the core by rotational mixing. 

\begin{figure}
    \centering
    \begin{minipage}{1\columnwidth}
    \hspace{-4.5mm}
    \includegraphics[width=1.05\linewidth,trim={0 0cm 0 0.14cm},clip]{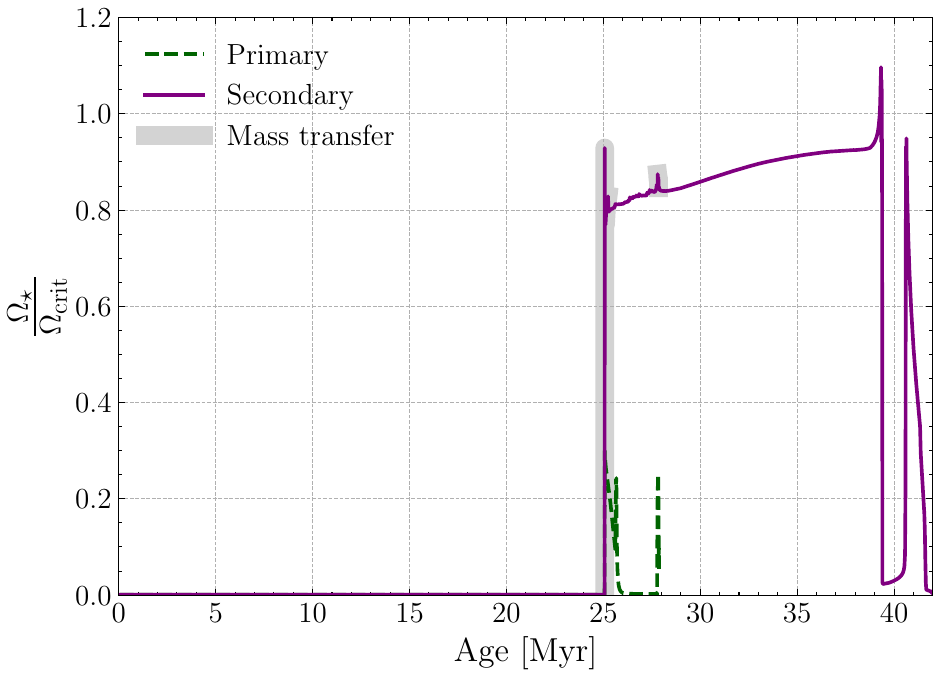}

    \end{minipage}
    \caption{The surface angular velocity relative to critical rotation over the lifetime of both stars, starting with zero initial spin. The primary experiences two spin-ups due to tidal interactions, whilst the secondary experiences a sustained spin-up due to the brief mass-transfer event (highlighted in grey). Towards the later stages of the secondary evolution the surface rotation fluctuates as changes in radius result in changes in moment of inertia, ending with a gradual spin down via stellar winds. }    
    \label{fig:Omega-Age_M10P450}
\end{figure}

In Fig.\,\ref{fig:Omega-Age_M10P450} we show the evolution of the angular velocity relative to critical $ \Omega_\star / \Omega_\mathrm{crit}$, following the angular momentum gain in the secondary. After a brief spin-up of the primary (dashed-green) via tides, mass-transfer (highlighted in grey) spins up the secondary. The secondary (solid-purple) reaches near $0.9\,\Omega_\mathrm{crit}$ (our adopted definition of critical rotation) and retains much of the spin after mass transfer. Towards later stages in the secondary's evolution, the secondary experiences further spin-up/down due to radial expansions and contractions, which alter the moment of inertia within the star. 

\begin{figure}
    \centering
    \begin{minipage}{1.0\columnwidth}
    \hspace{-3mm}
	\includegraphics[width=1.052\linewidth,trim={0 0.15cm 0.42cm 0.00cm},clip]{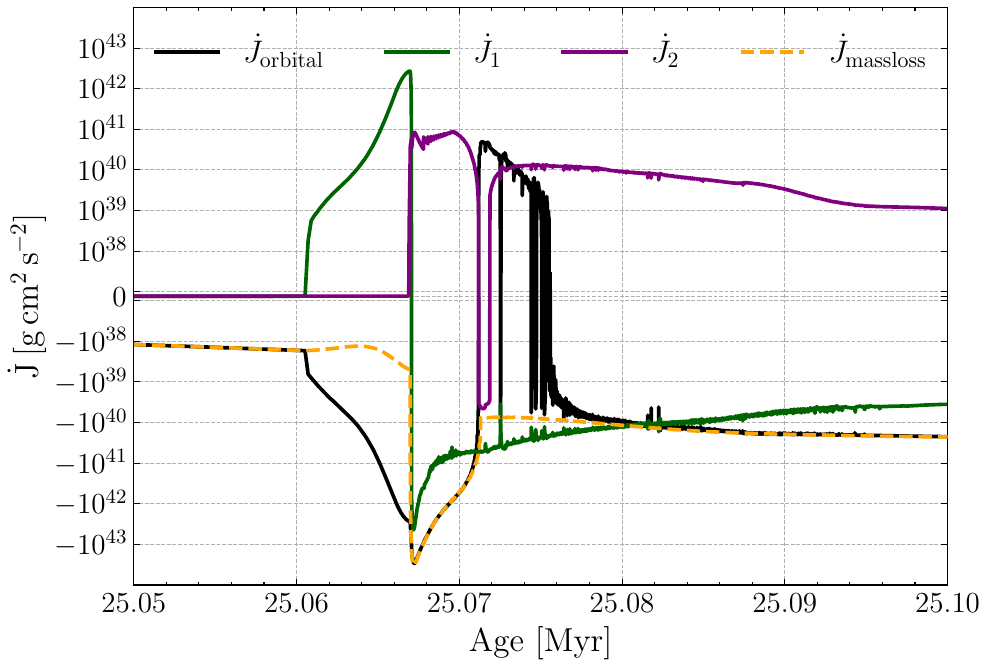}
    \end{minipage}
    \caption{The net torque during the brief period of mass-transfer on the orbital (black), primary (green) and secondary (purple), in addition to the total loss of angular momentum associated with mass loss (dotted yellow). Initially, both stars are losing mass due to stellar winds. Then the primary ($J_1$) removes angular momentum from $J_\mathrm{orb}$ via tides, resulting in $\dot J_\mathrm{orb}$ deviating from $\dot J_\mathrm{mass\, loss}$. Once mass-transfer begins, much of $J_{1}$, and some of $J_\mathrm{orb}$, is transferred to $J_{2}$ during the mass-transfer period. Once the surface reaches critical rotation, the disc mechanism activates, transferring angular momentum from $J_2$ to $J_\mathrm{orb}$ and demonstrating the disc's capacity to alter angular momentum exchange significantly. When not perturbed by tidal or disc interactions, $\dot J_\mathrm{orb}$ traces $\dot J_\mathrm{mass\, loss}$, including a boost in $\dot J_\mathrm{mass\, loss}$ due to thermally limited accretion removing unaccreted mass from the binary.}
    \label{fig:Jdot-Age_M10P450}  
\end{figure}

In Fig.\,\ref{fig:Jdot-Age_M10P450} we present $\dot J$ during the mass-transfer stage displayed using a symlog plot (allowing for both negative and positive values in log space). The mass loss driven component of $J_\mathrm{orb}$ (in yellow, labelled $\dot J_\mathrm{mass\, loss}$) remains negative and steady before and after binary interactions, driven by stellar winds. During the radial expansion of the primary, $\dot J_1$ becomes positive due to tidal interactions, during which $\dot J_\mathrm{orb}$ deviates from $\dot J_\mathrm{mass\, loss}$ as additional orbital angular momentum is removed and transferred onto $J_1$. Once mass-transfer initiates, $\dot J_1$ becomes negative due to transferring the tidally spun-up mass onto $\dot J_2$, which becomes positive. During this initial stage of mass transfer, before the surface is sufficiently spun up to initiate the disc mechanism, $\dot J_\mathrm{orb}$ again traces $\dot J_\mathrm{mass\, loss}$ (now dominated by loss of mass via $\beta_\mathrm{therm}$). Once the surface approaches critical rotation, the disc $\dot J$ is activated, resulting in brief bursts of positive $\dot J_\mathrm{orb}$. Depending on the magnitude of disc interactions $\dot J_2$ is either positive or negative. This depends on whether the angular momentum removed via the disc exceeds the net increase in total angular momentum capacity of $J_2$, as new mass is accreted. Towards the end of mass-transfer $\dot J_\mathrm{orb}$ realigns with $\dot J_\mathrm{mass\,loss}$ (from stellar winds).

\subsection{Mass-transfer efficiency} \label{subsec:MTE}

\begin{figure}
    \centering
    \begin{minipage}{1\columnwidth}
    \hspace{-4.5mm}
    \includegraphics[width=1.05\linewidth,trim={0 0 0 0.0cm},clip]{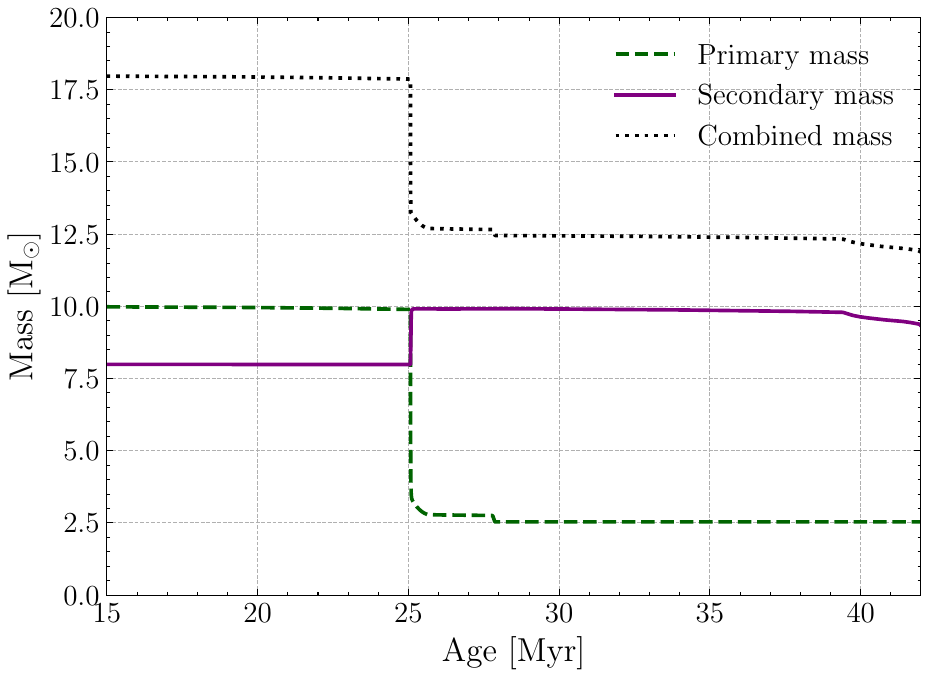}
    \end{minipage}
    \caption{The exchange of mass between different components of the binary during mass transfer. The primary transfers $\approx 70\,\%$ of its mass, of which $\beta_\mathrm{eff}\approx 30\,\%$ reaches the secondary, resulting in a net loss of $ \approx 27\,\%$ of the binary's total mass.}
    \label{fig:M-Age_M10P450}
\end{figure}

A primary motivation for incorporating our accretion disc model is to investigate an alternative to rotationally limited accretion in stellar models with differential rotation. Fig.\,\ref{fig:M-Age_M10P450} illustrates the mass evolution of the reference binary simulation and its components. During the brief but intensive mass-transfer episode, roughly $\beta_\mathrm{eff}\approx 30\,\%$ of the primary's transferred mass is retained by the secondary, with the remainder being lost due to the thermally limited accretion scheme. 

\begin{figure*}
    \centering    
    \begin{minipage}{0.48\textwidth}
    \hspace{-5mm}
    \includegraphics[width=1.08\linewidth,trim={0 0 0 0.0cm},clip]{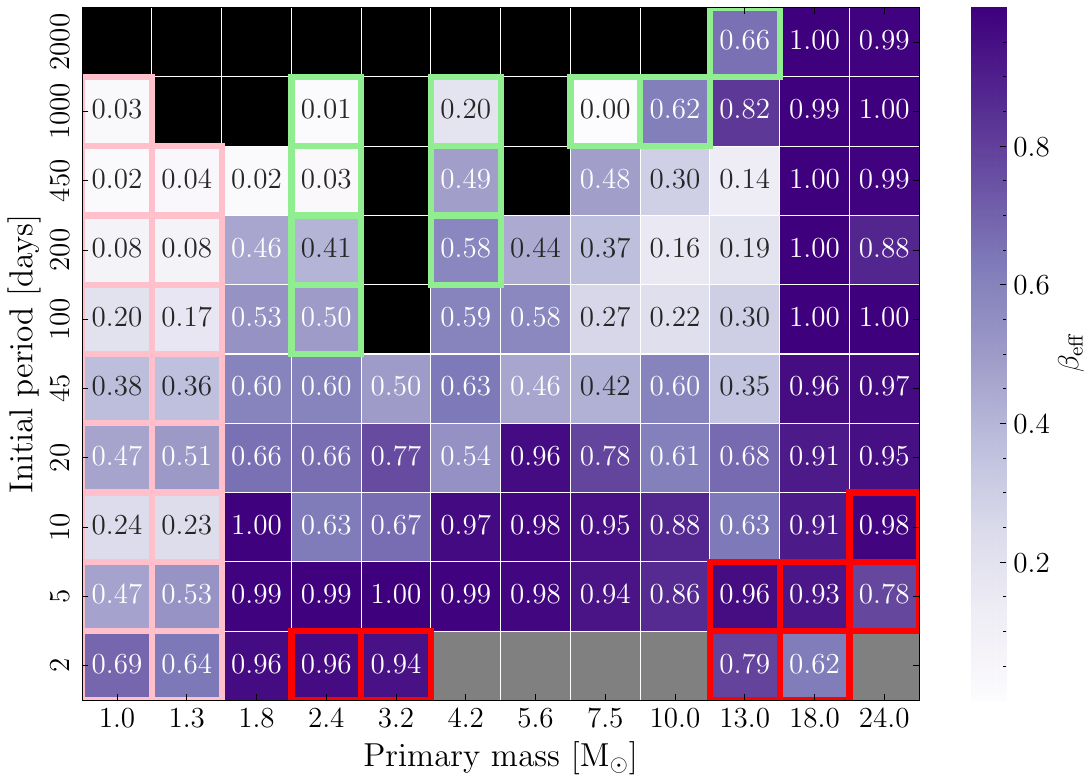}
    \end{minipage}
    \hfill
    \begin{minipage}{0.48\textwidth}
    \hspace{-2mm}
    \includegraphics[width=1.08\linewidth,trim={0 0 0 0.0cm},clip]{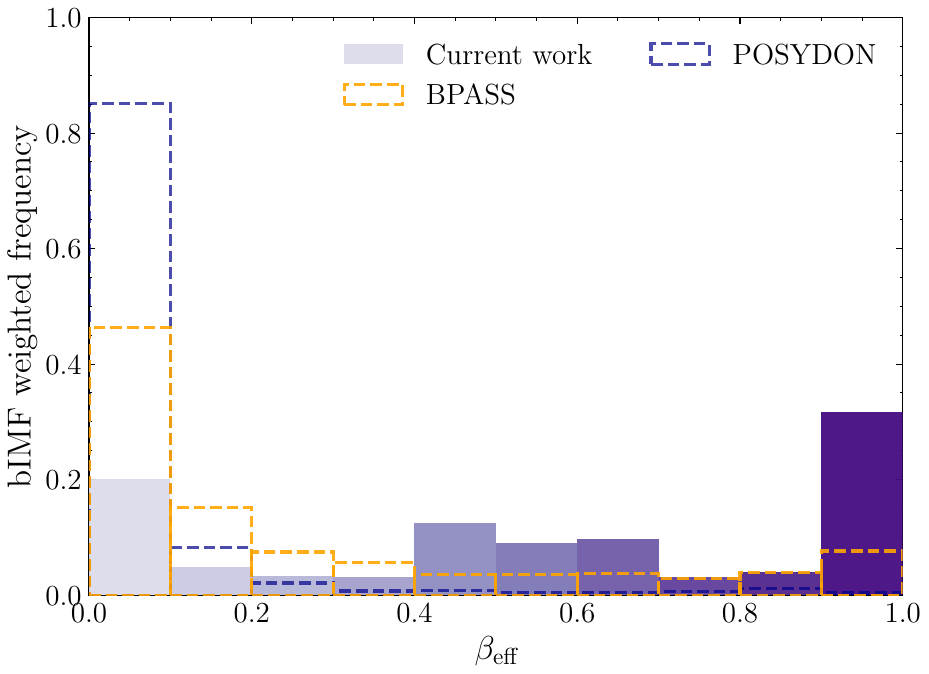}\vspace{-0.0cm}
    \end{minipage}
    \caption{\textit{Left:} A heatmap of $\beta_\mathrm{eff}$ for a binary mass ratio of 0.8 at Solar metallicity, across initial periods and primary masses. We only assign a $\beta_\mathrm{eff}$ if the model completes stable mass-transfer. The black squares represent cases where negligible mass-transfer occurs ($\dot{m}_\mathrm{negligible} < 10^{-10}\, \mathrm{M}_{\odot}\,\mathrm{yr}^{-1}$). The grey squares end, or undergo numerical instabilities, owing to the secondary filling its Roche lobe. We highlight models that undergo case A (red), case C (light green) mass transfer, or where the accretor has a convective envelope (light pink) at the start of mass transfer. \textit{Right:} Binary fraction and IMF weighted histogram for binaries within the same q= 0.8 parameter space as shown in the heatmap, comparing the disc model simulations with \textsc{bpass}, and \textsc{posydon} grids. The resulting distribution for the current work shows a flatter distribution, closer to that of \bpass than \textsc{posydon}. Distributions at SMC metallicity are presented in Appendix\,\ref{section:appendixSMC}.}
    \label{fig:beff_grid}
\end{figure*}

To obtain a broader view of the mass-transfer efficiency, we
calculate
\begin{equation*}
    \beta_\mathrm{eff}
= \frac{\Delta M_\mathrm{2,MT}}{\Delta M_\mathrm{1,MT}}\, ,
\end{equation*}
for binaries that complete stable mass transfer, by summing the mass lost and gained due to mass transfer (whilst excluding mass changes due to stellar winds). We present a $\beta_\mathrm{eff}$ heatmap for $q=0.8$ in the left panel of Fig.\,\ref{fig:beff_grid}. Across most of the parameter space, we find $\beta_\mathrm{eff}$ values ranging from approximately $0.4$ to $1.0$. We plot the corresponding $\beta_\mathrm{eff}$ distribution (shaded in purple) as a histogram in the right panel of Fig.\,\ref{fig:beff_grid}, where the disc model produces a relatively uniform distribution of $\beta_\mathrm{eff}$ values for $q=0.8$. Several regions of parameter space systematically depart from this overall behaviour.

We find low values of $\beta_\mathrm{eff}$ at lower masses, $M_1 \leq 1.3\,\mathrm{M}_\odot$, where mass is transferred onto an accretor with a convective envelope (highlighted pink), which may alter the thermal timescale of the accretor. A similar decrease in $\beta_\mathrm{eff}$ occurs at large separations, where mass transfer commences during helium-shell burning (highlighted light green). Here, the donor expands more rapidly than during hydrogen-shell burning, producing higher mass-transfer rates that exceed the thermal accretion limit by a larger factor.

At intermediate masses, there is a cluster of binaries, including our reference binary, that experience less efficient mass transfer around $M_1 \approx 10\,\mathrm{M}_\odot$ and $P \approx 200\,\mathrm{d}$. In contrast, for binaries with higher masses, $M_1 \geq 18\,\mathrm{M}_\odot$, we find a sharp transition to near-conservative mass transfer, suggesting a potential disagreement with the lower $\beta_\mathrm{eff}$ values inferred for observed WR+O binaries \citep{Nuijten2025}. To better understand this transition, we compare $\dot{m}_\mathrm{MT}$ with the accretor's $\dot{m}_\mathrm{thermal}$ for $13\,\mathrm{M}_\odot$ and $18\,\mathrm{M}_\odot$ primary masses at an initial period of $200\,\mathrm{d}$. The transition is only partially explained by differences in the pre-MT thermal limits and peak $\dot{m}_\mathrm{MT}$, both of which are within an order of magnitude for the two mass bins. During mass transfer, however, $\dot{m}_\mathrm{thermal}$ increases by more than an order of magnitude for the $18\,\mathrm{M}_\odot$ primary, allowing $\dot{m}_\mathrm{MT}<\dot{m}_\mathrm{thermal}$ to be maintained and resulting in near-conservative mass transfer. In contrast, at the onset of mass transfer in the $13\,\mathrm{M}_\odot$ system, $\dot{m}_\mathrm{thermal}$ lags behind $\dot{m}_\mathrm{MT}$. The mass-transfer rate therefore initially exceeds the thermal limit by almost an order of magnitude, resulting in a low $\beta_\mathrm{eff}=0.19$.

The $13\,\mathrm{M}_\odot$ example is representative of this region of low $\beta_\mathrm{eff}$ and illustrates a positive feedback loop. The initially intense mass transfer overwhelms the thermal limit, resulting in inefficient mass gain by the accretor. This delays the donor and accretor reaching a mass ratio of unity, extending the orbital-shrinking phase ($q<1$) and maintaining more intense mass transfer. The $18\,\mathrm{M}_\odot$ system avoids this feedback loop because the accretor's thermal timescale rapidly adjusts at the onset of mass transfer, allowing it to retain most of the transferred mass and rapidly reach $q>1$, after which mass transfer widens the orbit.

We find that very little mass is transferred in our $3.2\,\mathrm{M}_\odot$ bin at initial periods above $45\,\mathrm{d}$. This transition occurs close to the initial mass where the post-main-sequence expansion changes character. Compared to lower-mass stars, the donor undergoes less radial expansion, while wind mass loss widens the orbit sufficiently to prevent Roche-lobe overflow. At higher masses, the post-main-sequence expansion becomes much larger and overcomes the orbital widening, allowing Roche-lobe overflow during hydrogen-shell burning. This differs from the lower-mass bin (e.g. $2.4\,\mathrm{M}_\odot$), where the Roche lobe is filled during helium ignition, and the higher-mass bin (e.g. $4.2\,\mathrm{M}_\odot$), where it is filled during hydrogen-shell burning.

In red we highlight models undergoing case A mass transfer. Above $2.4\,\mathrm{M}_\odot$, mass-transfer efficiencies decrease with increasing mass, reproducing the trend found by \citet{Henneco2024} for rotationally limited case A mass transfer. We plot and further discuss case A mass-transfer efficiencies in Section\,\ref{subsec:CWB}. In grey we highlight case A mass-transfer events where the model ends due to contact or numerical instability whilst $R_2>0.99\,R_\mathrm{RL}$.

\subsection{Comparison with published binary grids}
\label{subsec:CWB}

We compare our model with other accretion limiting models using binary evolution codes, for the reference binary simulation ($10\, \mathrm{M}_\mathrm{\odot}$ + $8\, \mathrm{M}_{\odot}$ with $450\, \mathrm{d}$ initial orbital period) with identical parameters between \textsc{posydon} v2.0 \citep{Andrews2024}, \textsc{bpass} v2.2 \citep{BPASS2018}, and our model.

\textsc{posydon} includes rotation and thermally limited accretion within \textsc{mesa}, and thereby provides useful reference points for the current state of the art in binary evolution simulations employing \textsc{mesa}. \textsc{bpass} provides a useful comparison as it makes use of the thermally limited accretion scheme with limited modelling of rotational effects (hence no rotation limited accretion).

We compare the $\beta_\mathrm{eff}$ distributions, for $q=0.8$, of \textsc{bpass} (dashed yellow) and \textsc{posydon} (dashed blue) in the histogram of Fig.\,\ref{fig:beff_grid} (right panel), for the same initial primary mass and orbital period ranges at Solar metallicity. The \textsc{posydon} distribution is representative of rotationally limited accretion, being almost entirely concentrated around values of $< 10\,\%$. The \textsc{bpass} distribution reflects a thermally limited accretion sample that results in a more diffuse distribution biased towards lower values. In contrast, the disc model produces a more uniform distribution of $\beta_\mathrm{eff}$ for $q=0.8$, with minor bumps at either extreme. The inclusion of lower mass ratio bins (not shown here) shifts all grids to a bottom-heavy distribution. The fraction of binaries with $\beta_\mathrm{eff}>20\,\%$ for each given grid then becomes approximately; \posydon: $10\,\%$, \bpass: $20\,\%$ and for the current work: $35\,\%$ with a bump near $\beta_\mathrm{eff}=100\,\%$.

\begin{figure}%
\vspace{3mm}
%\centering
    \begin{minipage}{1\columnwidth}
    \hspace{-4.5mm}
	\includegraphics[width=1.05\linewidth,trim={0cm 0 0 0cm},clip]{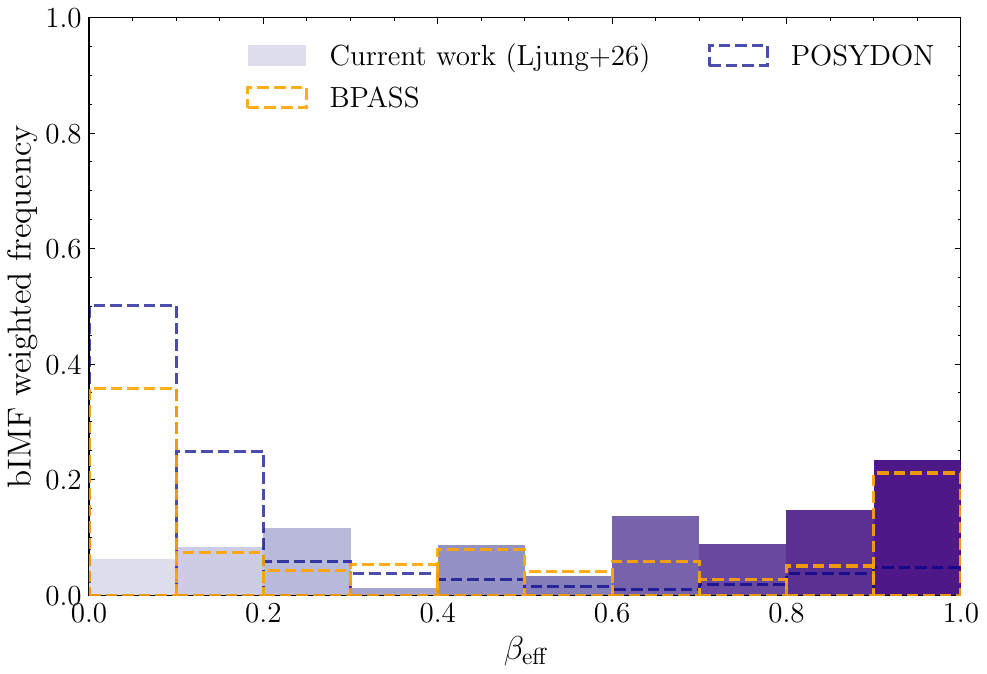}
    \end{minipage}

    \caption{$\beta_\mathrm{eff}$ distribution (weighted by binary fractions and IMF based on \citealt{Moe2017}) for case A mass transfers across mass ratio bins $0.1<q<0.9$, for \textsc{bpass}, \textsc{posydon}, and the disc model. Whilst \posydon predicts a bottom-heavy distribution and the disc model suggests a flat distribution, the \bpass distribution shows a mix of both. None of the models support the approximation that case A mass-transfer consistently results in $\beta_\mathrm{eff}\approx 100\,\%$.}
    \label{fig:b_eff_caseA}
    
\end{figure}
The primary source of disagreement between rotationally limited models and observations is for binaries with large initial separations that undergo case B mass transfer. In the case A regime, tidal interactions allow for partial dissipation of excess angular momentum into tides, which is thought to allow for larger mass-transfer efficiencies even in rotationally limited models \citep[e.g.][]{Shao2021}. We present our results, for all of our mass ratio bins, in Fig.\,\ref{fig:b_eff_caseA} to compare the behaviour of the differing models in the case A regime. The resulting $\beta_\mathrm{eff}$ distribution for \posydon is no longer limited to $\beta_\mathrm{eff} < 10\,\%$, instead forming a bottom-heavy distribution, as presented in \citet{Sen2022}. Despite the inclusion of lower mass ratio bins, the case A distribution for the disc model looks similar to the $q=0.8$ counterpart (which also includes case A). The distribution remains relatively flat for the disc model and results in a clearer bimodal distribution for \textsc{bpass}. This supports the understanding that tides can mitigate the impact of rotation driven accretion limits, but challenges the assumption that case A mass-transfer can be treated as conservative.

\begin{figure*}%
\centering
    \begin{minipage}{0.48\textwidth}
    \hspace{-4mm}
	\includegraphics[width=1.08\linewidth,trim={0 0 0 0cm},clip]{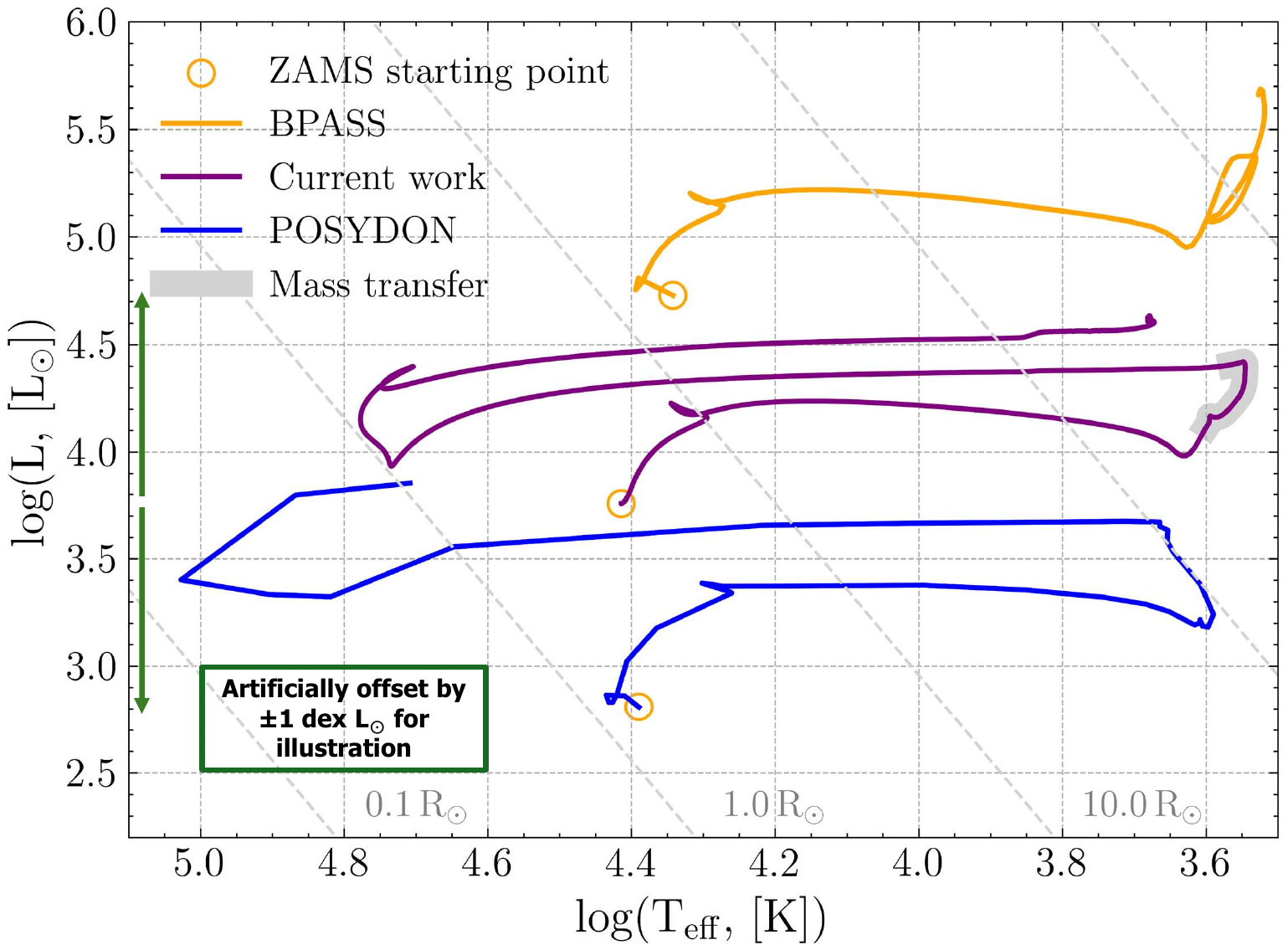}
    \end{minipage}
    \hfill
    \begin{minipage}{0.48\textwidth}
    \hspace{0mm}
	\includegraphics[width=1.08\linewidth,trim={0 0 0 0cm},clip]{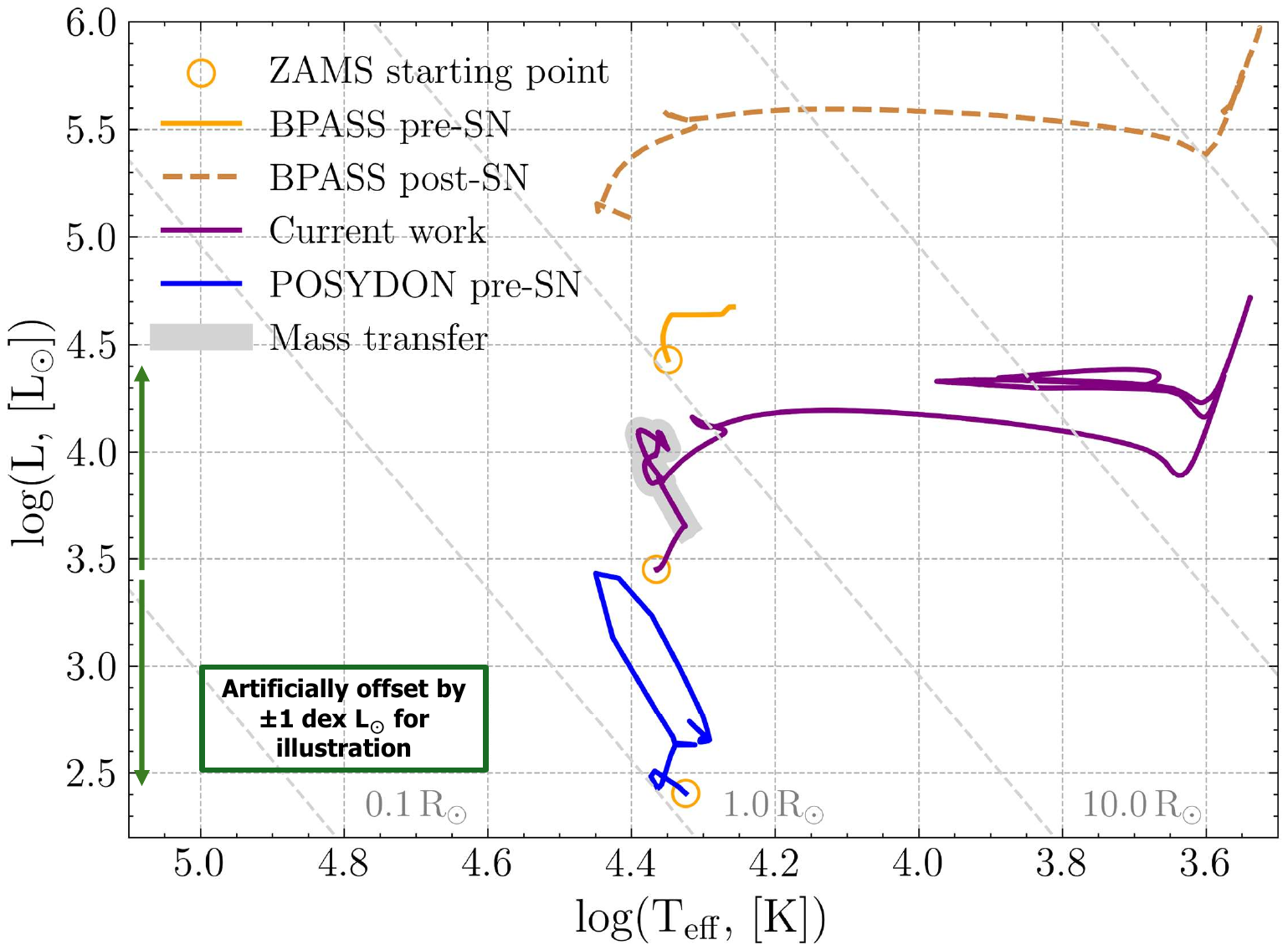}
    \end{minipage}
    \caption{HR diagrams of the primary (\textit{left}) and secondary (\textit{right}), comparing our reference binary simulations (purple, with mass-transfer highlighted in grey) against the \textsc{bpass}  (solid yellow and dashed yellow, offset by $+1\, \mathrm{dex}\, \mathrm{L}_{\odot}$) and \textsc{posydon} (blue, offset by $-1\, \mathrm{dex}\, \mathrm{L}_{\odot}$). The tracks for the primary are similar, showing that our model does not drastically alter its evolution. The \textsc{posydon} secondary track cuts off due to the simulation ending when the primary reaches the end of its evolution (post-SN not shown). The \textsc{bpass} secondary track is split into two, showing the initial evolution (\textsc{bpass} pre-SN, solid yellow) and the subsequent isolated evolution from a $13\, \mathrm{M}_{\odot}$ ZAMS (\textsc{bpass} post-SN, dashed yellow, no artificial offset from \textsc{bpass} pre-SN) once the \textsc{bpass} primary reaches core carbon depletion.}
    \label{fig:HR_comp}
\end{figure*}

In Fig.\,\ref{fig:HR_comp} we compare our reference binary system simulation (incorporating the disc model) with evolutionary tracks with the same initial conditions from the \textsc{bpass} and \textsc{posydon} grids. The left panel of Fig.\,\ref{fig:HR_comp} compares the evolution of the primary component in the binary (noting a vertical offset of $\pm\, 1\, \mathrm{dex}\, \mathrm{L}_{\odot}$ applied to \textsc{bpass} and \textsc{posydon} models for illustration purposes). The three models behave similarly until they reach the dip, around $\mathrm{log}(T_\mathrm{eff})=3.6$, associated with mass transfer. The \textsc{posydon} primary follows a similar track to our model, however, it ends at a different point in the Hertzsprung--Russell (HR) diagram, which we attribute to differences in post-RLOF wind prescriptions, where \textsc{posydon} makes use of \citet{Nugis2000} whilst we make use of \citet{Vink2017}, which retains small amounts of hydrogen post-RLOF \citep{Gilkis2019}. The \textsc{bpass} primary diverges from the other models as it bypasses the `blue loop' associated with intermediate-mass helium core burning. It is unclear why it is absent, but it might relate to helium flash modelling approximations in the underlying \textsc{stars} code (see \citealt*{Walmswell2015} for a detailed investigation of the blue loop phenomenon).

The HR diagram of the secondaries, in Fig.\,\ref{fig:HR_comp} (right panel), better illustrates the differences in behaviour and evolution between the models. The evolution of the \textsc{bpass} and \textsc{posydon} secondaries ends early as the model stops once the primary collapses. Both grids restart the simulation as a post-SN compact object + secondary, starting from main-sequence hydrogen or helium burning. In contrast, our models evolve the binary continuously through the primary's terminal core collapse, preserving the thermal, chemical and rotational state of the secondary. Restarting the secondary evolution is often a reasonable approximation as it approximates the effect of rotational mixing and introduction of new hydrogen fuel (by starting in the homogeneous pre-main-sequence state). When either grid restarts their secondaries post-SN, they only retain information about their mass, not the chemical, rotational, or thermal state of the secondary at the end of the first run. Owing to limited mass change, the \textsc{posydon} secondary would retain little information from the previous simulation, hence, we opt to include the \textsc{bpass} post-SN track but not the \textsc{posydon} one.

The \textsc{posydon} secondary (right panel) illustrates a loop around $\mathrm{log}(T_\mathrm{eff})=4.35$, also seen in our model, arising at the onset of mass-transfer and its associated thermal non-equilibrium state. This is missing in \textsc{bpass}, which during primary evolution (the first run) approximates the secondary evolution using analytic formulae from \citet{Hurley2002} which does not track their thermal state. This results in a post-SN \textsc{bpass} track with a $13\, \mathrm{M}_{\odot}$ secondary (compared to $8\, \mathrm{M}_{\odot}$ for \posydon and $10\, \mathrm{M}_{\odot}$ for our model) and a $1.4\, \mathrm{M}_{\odot}$ primary remnant ($3.0\, \mathrm{M}_{\odot}$ pre-SN), which highlights the uncertainty introduced by differing mass-transfer prescriptions. Whilst the \textsc{posydon} secondary luminosity remains steady, the \textsc{bpass} secondary luminosity increases by a factor $5$ after mass-transfer and our model results in an intermediate factor $2$ boost in luminosity. We note that the post-SN \textsc{bpass} track is not artificially offset from the pre-SN \textsc{bpass} track, rather, the discontinuity suggests that during primary evolution, the analytic model fails to capture the significant increase in secondary luminosity during and after mass-transfer. 

\begin{figure}%
\vspace{3mm}
%\centering
    \begin{minipage}{1\columnwidth}
    \hspace{-4.5mm}
	\includegraphics[width=1.05\linewidth,trim={0cm 0 0 0cm},clip]{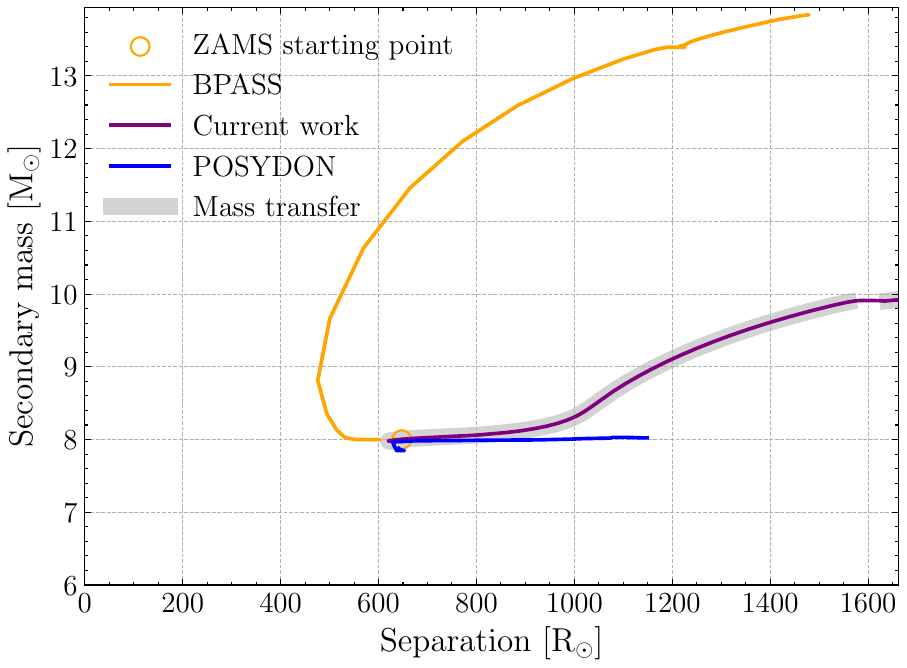}
    \end{minipage}

    \caption{Evolution of the secondary mass and binary separation, for the reference binary, between \textsc{bpass}, \textsc{posydon} and our model (mass-transfer highlighted in grey). \textsc{bpass} shows a steady accretion of mass, starting with a decreasing separation (whilst the mass ratio is $<1$) and later increasing separation (when the mass ratio becomes $>1$). The \textsc{posydon} model has limited mass gain due to the rotation limited accretion scheme. During mass transfer, our model transiently decreases in separation, but subsequently increases in mass and separation. The post-SN evolution of our model and the \textsc{bpass} models are not shown here.}
    \label{fig:M2-sep_P450_BPOSYDON}
    
\end{figure}

Whilst the HR diagrams illustrate differences in stellar evolution tracks, we plot the secondary mass against separation in Fig.\,\ref{fig:M2-sep_P450_BPOSYDON} to gain insight into the differences in their resulting compact object binaries. These are the parameters that best trace the initial mass and $J_\mathrm{orb}$ of compact binaries and are most affected by the disc and thermally/rotationally limited accretion models (as the primary will transfer/lose similar amounts of mass for different accretion limiting models). Starting at the yellow circle, for the disc model, mass-transfer (highlighted in grey) first results in the separation briefly decreasing, due to a combination of mass-transfer onto the less massive secondary ($q<1$) and angular momentum transfer from $J_\mathrm{orb}$ to $J_2$. This is followed by an increase in separation from a combination of mass-transfer onto a more massive secondary ($q>1$), disc transfer of angular momentum to $J_\mathrm{orb}$, and mass removed from the binary via the thermally limited accretion scheme. After mass-transfer the secondary experiences a gradual loss of mass via winds, which leads to a slight increase in separation. At the end of the primary's life, the binary separation jumps due to the SN kick, extending past the maximal separation in the plot.

Due to the rotationally limited accretion scheme in the \textsc{posydon} model, its secondary will accrete only about $2\,\%$ of the transferred mass, resulting in limited mass change, but a gradual increase in separation due to non-conservative mass-transfer and stellar winds. In contrast, the \textsc{bpass} model only applies thermally limited accretion, limiting accretion to $10\times\, \dot m_\mathrm{therm}$ (as opposed to the $1\times\, \dot m_\mathrm{therm}$ used in our simulations), thereby most of the transferred mass is accreted ($\beta_\mathrm{eff} \approx 67\,\%$, compared to our $\beta_\mathrm{eff} \approx 30\,\%$). In the absence of the disc-driven $\dot J$, and a more lenient $\beta_\mathrm{therm}$, as the \textsc{bpass} secondary accretes mass, it initially (whilst $q < 1$) spirals inwards more than our model. Once the mass fraction passes $q=1$, the separation instead increases. Once the primary reaches the end of its evolution, the \textsc{bpass} SN prescription predicts that the binary gets disrupted (we do not display the post-SN \textsc{bpass} secondary in this plot). In this case, as our model made use of a more conservative thermally limited accretion prescription (with $1\times\, \dot m_\mathrm{therm}$) than \textsc{bpass}, our secondary ends up with a significantly lower mass.

\section{Discussion}
\label{sec:discussdiscs}

\subsection{Implications for mass-transfer efficiency models} \label{subsec:IFM}

\begin{figure*}% 
    \begin{minipage}{0.7\textwidth}
    %\hspace{-3mm}
	\includegraphics[width=1.00\linewidth,trim={0 0 0 0.0cm},clip]{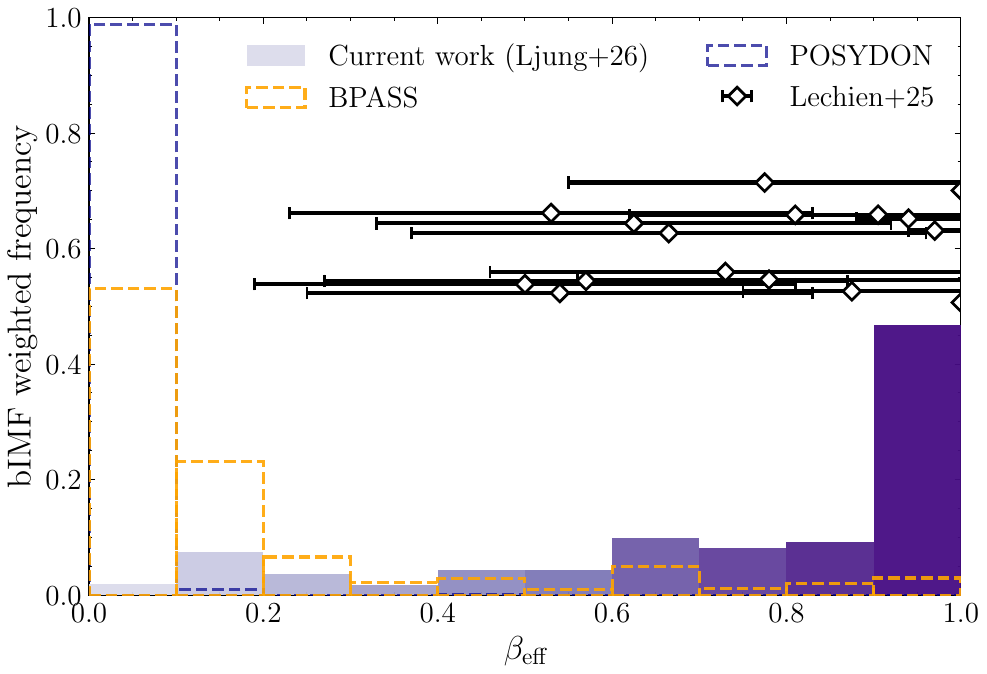}
    \end{minipage}
    \caption{Distribution of $\beta_\mathrm{eff}$ for \textsc{posydon}, \textsc{bpass} (without the rotational constraint), and the disc model, limited to the same parameter space as the observed binaries in \citetalias{Lechien2025}: $0.2< M_{1,\,\mathrm{post-MT}}<3\,\mathrm{M}_{\odot}$, $3< M_{2,\,\mathrm{post-MT}}<20\,\mathrm{M}_{\odot}$, $10<P_\mathrm{orb, \,post-MT}<400\,\mathrm{d}$, $0.5\leq q$, $10000< T_{\mathrm{eff}, 2}<40000\,\mathrm{K}$, $0.7\leq \frac{\Omega_\mathrm{surface}}{\Omega_{\mathrm{crit}}}$ (the rotational constraint is not applied to \bpass models). Weighted according to binary fractions and IMF from \citet{Moe2017}. The `best guess' ranges of $\beta_\mathrm{eff}$ according to \citetalias{Lechien2025} are presented as a collection of black diamonds. The \textsc{bpass} dataset shows partial overlap with \citetalias{Lechien2025}, whilst the current work better reproduces the data for the given assumptions.}
    \label{fig:beff_obs}
\end{figure*}

A central motivation of this work is to address the tension between rotationally limited accretion models and observationally inferred mass-transfer efficiencies in interacting binaries. In Section\, \ref{subsec:CWB} we compared the effective mass-transfer efficiency, $\beta_\mathrm{eff}$, obtained with our disc-mediated angular momentum transport model to predictions from \textsc{bpass} and \textsc{posydon}, which represent various combinations of thermally limited and rotationally limited prescriptions.

The resulting $\beta_\mathrm{eff}$ distributions highlight a qualitative difference between models. Rotationally limited accretion models strongly suppress mass gain once the secondary approaches critical rotation, producing distributions sharply peaked at low $\beta_\mathrm{eff}$, consistent with long-standing theoretical expectations that only a small amount of accreted mass is required to spin a star to near-critical rotation. Thermally limited accretion allows substantially higher mass gain, but still favours relatively low efficiencies. We note that the distribution presented by \citetalias{Lechien2025} (not shown) exhibits an isolated spike at $\beta_\mathrm{eff}\simeq 1$ that is not reproduced by the corresponding \textsc{bpass} models. 

To compare our results with observations, we apply IMF and binary distribution weightings based on observations (e.g. \citealt{Moe2017}) and restrict the disc, \textsc{bpass}, and \textsc{posydon} models to the same parameter space as the sdOB+Be binaries analysed by \citetalias{Lechien2025}. Specifically, $0.2< M_{1,\,\mathrm{post-MT}}<3\,\mathrm{M}_{\odot}$, $3< M_{2,\,\mathrm{post-MT}}<20\,\mathrm{M}_{\odot}$, $10<P_\mathrm{orb, \,post-MT}<400\,\mathrm{d}$, $10000< T_{\mathrm{eff}, \,2}<40000\,\mathrm{K}$, $0.7\leq \frac{\Omega_\mathrm{surface}}{\Omega_{\mathrm{crit}}}$, but allowing for mass ratios down to their assumed limit of $\mathrm{q}\geq 0.5$. As the observed binaries were observed to have completed and survived their mass-transfer phase, we remove any modelled binaries that do not survive past mass transfer. Furthermore, we sample the $\beta_\mathrm{eff}$ of the simulated binaries at a point before they become compact objects, as the observed sample of sdOB+Be binaries have not yet formed compact objects.

In Fig.\,\ref{fig:beff_obs} we plot the resulting distributions for each grid based on the aforementioned criteria, along with the `best guess' ranges of $\beta_\mathrm{eff}$ according to \citetalias{Lechien2025} as black diamonds. The observed $\beta_\mathrm{eff}$ estimates heavily cluster towards higher values, with mid-points in the range $0.5<\beta_\mathrm{eff}<1.0$. We find that the disc model best reproduces the observed range, with a notable contribution to the $90-100\,\%$ bin from case A mass transfer events. The rotationally limited models are in strong disagreement. The filtered \textsc{bpass} distribution displays some overlap with the observed samples, but maintains a bottom-heavy distribution that is in slight tension with the data. 

These results are in agreement with the findings of \citetalias{Xing2026}, who implement a disc model inspired by \citetalias{Popham1991} and find a good match in Be accretor to helium donor masses and final periods, when compared to sdOB+Be binary observations. Their results favour a $\beta_\mathrm{eff}$ distribution centred about $50\,\%$ compared to the flat distribution of our model. However, it remains unclear whether such differences arise due to differing models or differences in sampled parameter space. This provides tentative but compelling evidence that disc-mediated angular momentum transport, or a physically equivalent mechanism, may offer a resolution to the mass-gain problem in rotating accretors.

Crucially, these agreements are achieved while modelling differential rotation. The ability of the disc to apply a negative torque once the stellar surface approaches critical rotation allows continued mass accretion without invoking extreme mass loss through winds, and their associated numerical instabilities, or artificially suppressing accretion, or rotation, altogether. Angular momentum exchange between a disc and a rapidly rotating star has long been recognised as a viable regulatory mechanism in other astrophysical contexts (e.g. \citealt*{Krticka2011}; \citealt{Haemmerle2017}). In this sense, the disc model decouples the mass and angular momentum budgets in a way that is difficult to achieve with wind-based prescriptions alone.

\subsection{Comparison to literature post-interaction mass-transfer efficiencies}\label{subsec:CLE}

\begin{figure*}
    \centering    
    \begin{minipage}{0.48\textwidth}
    \hspace{-5mm}
    \includegraphics[width=1.05\linewidth,trim={0 0 0 0.0cm},clip]{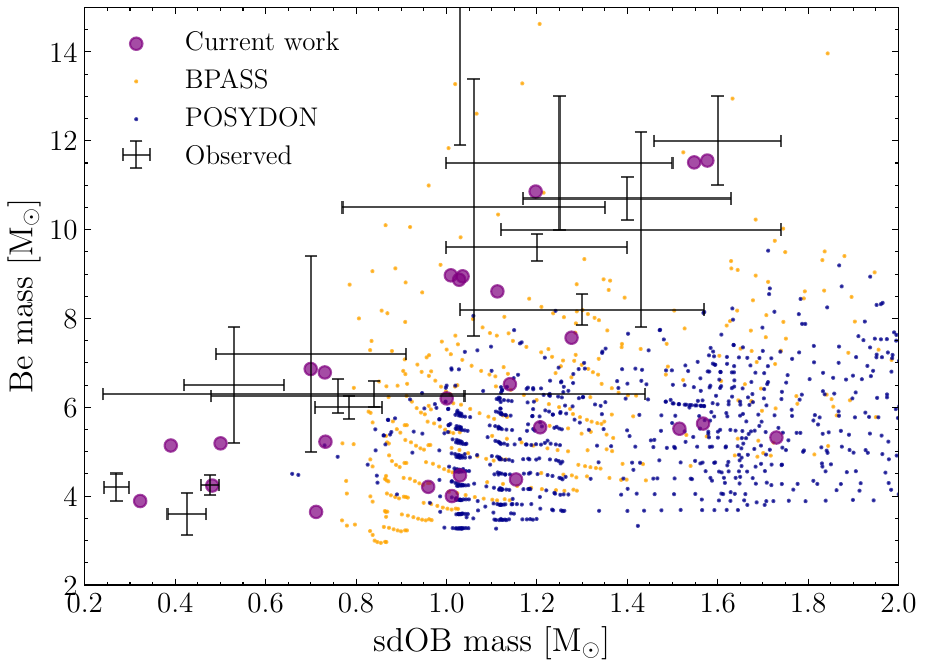}
    \end{minipage}
    \hfill
    \begin{minipage}{0.48\textwidth}
    \hspace{-2mm}
    \includegraphics[width=1.05\linewidth,trim={0 0 0 0.0cm},clip]{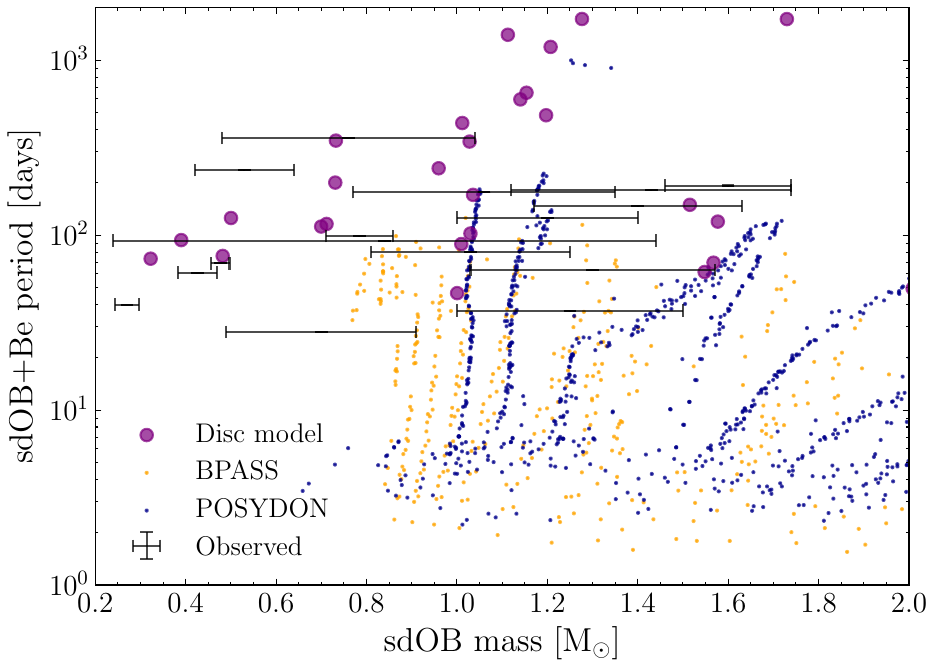}\vspace{-0.0cm}
    \end{minipage}
    \caption{The post-MT properties of sdOB ($10000< T_{\mathrm{eff}, 1}$, $\mathrm{surface\_hydrogen}<0.5$) + Be ($10000< T_{\mathrm{eff}, 2}<40000\,\mathrm{K}$, $0.7< {\Omega_\mathrm{surface}} / {\Omega_{\mathrm{crit}}}$) binaries for \textsc{bpass}, \textsc{posydon}, and the disc model. These are compared to the estimated parameters of the observed binaries in \citetalias{Lechien2025}. \textit{Left:} The post-MT Be against sdOB masses. \posydon is in tension with observations, whilst \bpass shows partial overlap with the distribution, however both fail to reproduce the lower end of the sdOB mass distribution, $ < 0.6\,\mathrm{M}_\odot$. The disc model better reproduces observations across the observed parameter space. All models produce a population of binaries with lower relative Be masses than the observed sample. \textit{Right:} The post-MT periods against sdOB masses. All models overlap with the observed period range, however, \bpass and \posydon predict a non-observed population of sdOB+Be binaries with periods $ < 20\,\mathrm{d}$, whilst the disc model predicts binaries at larger separations (which remain observationally unconstrained). SMC metallicity distributions are presented in Appendix\,\ref{section:appendixSMC}.}      
    \label{fig:BeHe}
\end{figure*}

Existing literature that makes use of sdOB + Be binaries as indicators of mass-transfer efficiency frequently compares the masses of both components to assess how well models reproduce observations \citep{Shao2014, Lechien2025, Xing2026}. We reproduce this comparison in the left panel of Fig.\,\ref{fig:BeHe}, which shows good agreement between the disc model and the observed distributions. Although \bpass shows partial overlap with the observed distribution, both \bpass and \posydon fail to reproduce sdOB masses below $0.6\,\mathrm{M}_\odot$. We find that for the disc model the $<0.6\,\mathrm{M}_\odot$ sdOB masses primarily arise from case A mass transfer events, whilst \bpass and \posydon fail to reproduce these masses, even with case A mass transfer. All models predict a population of Be companions with relative masses lower than those observed.

\citetalias{Xing2026} find that better overlap between their disc model and the observed population can be achieved by decreasing their mass dependent overshooting parameter, based on the implementation in \citet{Hastings2021}. In the relevant mass regime, this happens to bring their overshooting closer to what was used in this paper. However, this is in tension with recent observational constraints on overshooting by \citet{Lechien2026}, who propose a mass-dependent overshooting prescription that is notably higher than that used in this implementation and that of \citet{Hastings2021}, in the relevant mass regime.

Whilst sdOB-Be binaries provided evidence for the need of more efficient mass-transfer efficiencies, existing literature on Be-NS/X-ray binaries provide additional constraints that we can compare our model to. \citet{Shao2014} compare the resulting Be-NS binaries from rotationally limited, with $\beta_\mathrm{eff}=0.5$, and thermally limited models to observations, finding that $\beta_\mathrm{eff}=0.5$ is the only prescription that reproduces the observed secondary mass cut-off at $8\,\mathrm{M}_\odot$. They find that thermally limited models result in too high a lower-mass cut-off ($ > 13\,\mathrm{M}_\odot$), with the inverse problem for rotationally limited models ($ > 3\,\mathrm{M}_{\odot}$). We filter our grid to identify NS-Be stars, where $T_\mathrm{eff}>10^4\, \mathrm{K}$ and the surface is $>0.7\,\Omega_\mathrm{crit}$. We opt not to filter based on separation as we do not sample variations in SN kicks and significant uncertainties remain in the orbital angular momentum evolution. Our distribution of masses suggests a lower cut-off value around $>6.7\,\mathrm{M}_\odot$, suggesting a closer alignment with observations compared to thermal and rotational limits. When extending this comparison to Small Magellanic Cloud (SMC) metallicity Be-NS binaries, observations in \citet{Coe2015} suggest a lower mass limit of $6$--$7\, \mathrm{M}_\odot$ (based on the absence of spectral classes below B2e), which overlaps with our $>6.4\, \mathrm{M}_\odot$ cut-off for SMC metallicity.

Comparisons with the SMC Be X-ray binary population by \citet{Vinciguerra2020}, which test a range of case B mass-transfer efficiencies (conservative, thermally limited, $\beta_\mathrm{eff}=0.5$, $\beta_\mathrm{eff}=0.75$, and $\beta_\mathrm{eff}=0.0$), suggest that thermally limited accretion is inconsistent with the observations. Their thermally limited models predict a distribution strongly peaked at low mass-transfer efficiencies ($\beta_\mathrm{eff}\lesssim0.2$), in agreement with the \bpass\ results, whilst their best-fitting models adopt $\beta_\mathrm{eff}=0.5$. Their $\beta_\mathrm{eff}=0.75$ models also provide a good match to the observed population. Our model at SMC metallicity (Appendix\,\ref{section:appendixSMC}) instead predicts a flat distribution, which is in better agreement with their overall preferred range of $0.5\lesssim\beta_\mathrm{eff}<1$, than that of \bpass or \textsc{posydon}.

Observations suggest a distinction in the predicted mass-transfer efficiencies for higher mass binaries, like WR+O stars, where the literature suggests lower $\beta_\mathrm{eff}$ values than for Be-NS populations \citep{Petrovic2005, Shao2016, Schurmann2025, Nuijten2025}. This is in clear disagreement with our near-conservative mass-transfer for primary masses $\geq 18\,\mathrm{M}_\odot$ undergoing case B mass transfer. In conjunction with Oe stars being less prevalent than Be stars ($\approx4\,\%$ and $\approx30\,\%$ in the MW, respectively), in comparison with their no-disc counterparts \citep{Zorec1997, Negueruela2004}, this may hint at a mechanism by which massive accretors disrupt the formation or angular momentum transport of accretion and/or decretion discs, reverting to rotationally limited accretion. Further work is needed to determine potential disruption mechanisms (e.g. strong winds) and provide robust observational constraints.
    
\subsection{Other factors in assessing mass-transfer efficiencies} \label{subsec:OFI}

While the disc mechanism substantially alleviates rotational constraints, the overall mass-transfer efficiency remains sensitive to other poorly constrained aspects of binary evolution modelling. In particular, uncertainties associated with thermally limited accretion prescriptions remain significant. Factors like the frequent triggering of common envelopes in \textsc{bpass} may also bias $\beta_\mathrm{eff}$ values, sampled from surviving binaries. In our reference system, the post-mass-transfer secondary mass differs by several solar masses between models adopting $1\times\, \dot m_\mathrm{therm}$ and $10\times\, \dot m_\mathrm{therm}$, underscoring the potential systematic impact of this choice.

\citet{Wang2026} showed that, within rotation-limited accretion models, the effective mass-transfer efficiency can increase during the later stages of thermal-timescale mass transfer. As the accretor gains mass, its thermal timescale decreases, allowing it to return progressively towards thermal equilibrium. The resulting contraction increases the critical rotation rate, reducing the ratio $\Omega_\star/ \Omega_\mathrm{crit}$ and thereby permitting further accretion while remaining below the critical limit. Since both our models and \posydon\ employ \mesa, we likewise follow the thermal disequilibrium and structural response of the accretor self-consistently. However, we find that this effect leads to only a modest increase in the effective mass-transfer efficiency.

Recent work suggests that the maximum stable accretion rate before thermal disequilibrium and radial expansion set in is mass-dependent \citep{Lau2024,Shurmann2024}. This behaviour builds on earlier studies showing that accretion at rates exceeding the thermal adjustment timescale leads to envelope expansion and departure from equilibrium \citep[e.g.][]{Kippenhahn1977,Neo1977,Pols1994}. Lower-mass accretors appear capable of tolerating orders of magnitude higher multiples of $\dot m_\mathrm{therm}$ than more massive stars, which are more prone to envelope expansion. In this context, we interpret the onset of strong radial expansion as indicating that recently accreted material remains only weakly bound to the stellar envelope, and is therefore susceptible to subsequent removal through stellar winds or dynamical interactions. The presence of Roche-lobe constraints and binary torques is expected to further reduce the amount of mass required to trigger such expansion, supporting a conservative interpretation of the thermal stability limits inferred from single-star calculations. This implies that a single global multiple of $\dot m_\mathrm{therm}$ is unlikely to be optimal across the full stellar mass range. Implementing a mass-dependent thermal accretion limit could therefore both improve physical realism and allow the disc mechanism to activate earlier in systems with initially extreme mass ratios, potentially reducing the overproduction of contact and common-envelope events noted in previous studies.

A brief caveat concerns the adopted lower mass-ratio limit. In \citetalias{Lechien2025}, and in our comparison for Fig.\,\ref{fig:beff_obs}, a lower bound of $\mathrm{q}\geq0.5$ is assumed, motivated by the expectation that systems with smaller mass ratios generally undergo unstable mass transfer. However, we find that the disc model enables stable mass-transfer over a broader range of initial mass ratios, in most mass bins extending down to $\mathrm{q}=0.4$. Relaxing the $\mathrm{q}\geq0.5$ constraint would lower the inferred $\beta_\mathrm{eff,\,min}$ and shift the predicted $\beta_\mathrm{eff,\,obs}$ distribution in \citetalias{Lechien2025} towards less top-heavy values. Indeed, when including systems with $\mathrm{q}<0.5$, we identify additional binaries in the disc model that undergo stable mass-transfer and match the observed post-interaction parameter space, but exhibit very small $\beta_\mathrm{eff}$. The interpretation of these systems is not straightforward, particularly in light of literature suggesting that lower-mass accretors can sustain accretion rates several orders of magnitude above $\dot m_\mathrm{therm}$. Because implementing a mass-dependent $\dot m_\mathrm{therm}$ prescription lies beyond the scope of this work, we retain the $\mathrm{q}\geq0.5$ assumption of \citetalias{Lechien2025} for consistency in our observational comparison.

The fate of mass that is not immediately accreted also has important consequences for angular momentum evolution. If excess material is expelled via stellar winds, angular momentum is efficiently removed from the system. Alternatively, if mass is retained in a circumbinary configuration or ejected through three-body interactions, dynamical friction may drive inspiral in a manner analogous to a common envelope \citep*[e.g.][]{Soberman1997}. 

When comparing the period against sdOB mass distributions for sdOB+Be binaries, shown in the right panel of Fig. \ref{fig:BeHe}, both \bpass and \posydon predict large populations below the observed period distribution, which suggests a notable tension. Whilst upper limits on the periods remain poorly constrained due to observational biases against large period orbits, the same is not true for short periods which are more easily observed. 

The disc mechanism introduced here results in larger separations by providing a counteracting mechanism against inspiral that returns some angular momentum to the orbit. However, this introduces a different unconstrained parameter, the fraction of angular momentum that is transferred to the orbit, as opposed to lost through the $L_2$ or the outer disc. Three different implementations (fractions) have been explored thus far; the current implementation ($100\,\%$), \citetalias{Xing2026} ($50\,\%$), and \citetalias{Xing2026} ($0\,\%$). The current work ($100\,\%$) and \citetalias{Xing2026} ($50\,\%$) overlap with the observed periods, whilst returning $0\,\%$ of the angular momentum to the orbit is disfavoured. 

It is important to note that there are significant degeneracies in the proposed mechanisms (e.g. accretion and circumbinary discs, stellar winds) for regulating angular momentum in 1D stellar binary models, which will be difficult to disentangle. Hydrodynamic mass-transfer simulations \citep[e.g.][]{Dickson2024} will be needed for detailed tracking of angular momentum transport and loss within the system.

Stellar wind accretion prior to Roche-lobe overflow represents another secondary effect that may modestly increase $\beta_\mathrm{eff}$, particularly in systems with extreme mass ratios or long pre-interaction lifetimes. In our reference system, including wind mass-transfer for both components using the \textsc{mesa} \texttt{do\_wind\_mass\_transfer} Bondi--Hoyle prescription \citep{Bondi1944,Edgar2004} increases the effective mass-transfer efficiency by about two per cent. While this difference is minor in this specific case, wind accretion may play a more significant role in wider binaries or systems with more extreme mass ratios, where cumulative wind mass exchange may appreciably modify the thermal response of the accretor.

\subsection{Current modelling challenges and next steps} \label{subsec:CCW}

The dominant uncertainties in the present work stem from the simplified nature of the accretion disc prescription. For computational tractability, we adopt an analytic, steady-state disc model valid in the regime where deviations from Keplerian rotation are small. Although this approximation is supported by existing numerical studies and by order-of-magnitude checks, its domain of validity across stellar masses, accretion rates, and radiative regimes remains to be established.

Several physical ingredients are also neglected, including the outer disc boundary condition, despite likely truncation by the donor star and interaction with the incoming mass-transfer stream. Radiative pressure, disc warping, and time-dependent behaviour may all become important during rapid mass-transfer episodes. Ultimately, comparison with multi-dimensional hydrodynamic simulations will be required to assess whether the assumed negative torque can be sustained under more realistic conditions. Furthermore, we assume that the removed angular momentum is transferred to the orbital angular momentum, $\mathrm{J}_\mathrm{orb}$, through torques between the outer disc and the donor star. Although this assumption introduces additional uncertainty, it provides a testable hypothesis for the role of disc-based angular momentum exchange in interacting binaries.

Beyond the disc model itself, challenges remain in modelling super-critical rotation outside of mass transfer. Our use of decretion-like discs as a temporary angular momentum transport mechanism is motivated by numerical challenges with critical rotation outside mass-transfer and observations of Be stars and their circumstellar discs \citep*{Rivinius2013}, but remains heuristic. Similarly, resolving the shortest dynamical timescales in systems with extreme mass ratios continues to strain one-dimensional stellar evolution solvers, making it difficult to unambiguously distinguish physical instability from numerical failure \citep[e.g.][]{Pavlovskii2015,Marchant2021}.

The natural next steps are population-level comparisons with observed stellar clusters and compact object binaries, as well as direct contrasts with existing population synthesis frameworks. This will be essential for assessing the broader implications of disc-mediated mass-transfer for the formation of Be stars, X-ray binaries, and merging compact objects.

\section{Conclusion}\label{sec:conc}

We have introduced and implemented a disc-mediated angular momentum transport prescription for mass-accreting stars in interacting binaries, motivated by the star--disc boundary-layer model of \citet{Paczynski1991}. This prescription allows excess angular momentum to be transported away from the accretor as it approaches or is perturbed above critical rotation, enabling continued mass accretion without requiring extreme mass loss or strong magnetic coupling. Implemented within \textsc{mesa}, the model enables stable evolutionary calculations of differentially rotating binaries across extended mass-transfer phases.

Our results show that disc-mediated angular momentum redistribution can significantly alter the effective mass-transfer efficiency relative to commonly adopted thermally or rotationally limited prescriptions. In representative systems, accretors can gain several solar masses while remaining near, but below, critical rotation. This behaviour is broadly consistent with constraints from observed post-interaction binaries and helps alleviate tensions between theoretical models and
observations. At the same time, the prescription remains simple and computationally inexpensive, making it suitable for large grids of detailed stellar evolution models.

The present treatment is intentionally idealised and should be viewed as a first step towards more physically grounded modelling of star--disc coupling in binary evolution. Future work should explore the sensitivity to disc viscosity, boundary-layer structure, and coupling to orbital angular momentum, as well as comparisons with multidimensional simulations. Incorporating such processes more realistically has the potential to improve predictions for binary interaction outcomes, supernova progenitors, and compact-object populations.

The key findings of this work are as follows.
\begin{itemize}
\item A simple disc-mediated angular momentum extraction mechanism allows sustained accretion even near critical rotation.
\item Effective mass-transfer efficiencies can be substantially higher than in rotationally limited prescriptions.
\item The resulting accretor properties are compatible with observed post-interaction systems that have experienced significant mass gain.
\item The prescription is numerically stable and readily implementable in stellar evolution calculations.
\item Disc-mediated angular momentum transport may represent a missing ingredient in standard binary evolution models.
\end{itemize}

Overall, these results highlight that the treatment of angular momentum during mass-transfer remains a key uncertainty in binary evolution, and that even simple physically motivated prescriptions can lead to qualitatively different outcomes. Continued progress will require closer interplay between detailed stellar modelling, disc physics, and observational constraints.

\section*{Acknowledgements}

We thank the referee for a constructive review of the paper.
We thank Christopher Tout and Roman Rafikov for fruitful conversations on the physical properties and derivation of the disc model. We also thank Jan Henneco for sharing their results. We thank Max Briel for providing insights into the data processing for \textsc{bpass}.

%%%%%%%%%%%%%%%%%%%%%%%%%%%%%%%%%%%%%%%%%%%%%%%%%%
\section*{Data Availability}

The data underlying this article will be made available upon publication.

%%%%%%%%%%%%%%%%%%%% REFERENCES %%%%%%%%%%%%%%%%%%

% The best way to enter references is to use BibTeX:

\bibliographystyle{mnras}
\input{stardisc.bbl}

%%%%%%%%%%%%%%%%%%%%%%%%%%%%%%%%%%%%%%%%%%%%%%%%%%

%%%%%%%%%%%%%%%%% APPENDICES %%%%%%%%%%%%%%%%%%%%%

%\newpage

\appendix

\section{Derivation of disc equation}
\label{section:appendixEquationDerivation}

Our aim is to derive a simple accretion-disc prescription that can be coupled readily to \textsc{mesa} binary calculations, while retaining the key behaviour found numerically by \citetalias{Paczynski1991} and \citetalias{Popham1991} -- that near critical stellar rotation the disc can exert a negative torque even while the mass inflow rate remains positive. Such a mechanism could allow continued accretion without requiring strong magnetic coupling or excessive angular-momentum loss in winds.

The quantity we ultimately want is a relation between the stellar surface rotation $\Omega_\star$ and the angular-momentum flux through the inner disc, ${\dot J}$. We follow standard accretion disc formulation \citep[e.g.][]{Pringle1981}, beginning from the equation for conservation of angular momentum,
\begin{equation}
\frac{\partial \Omega}{\partial r}
=
-\frac{\dot m\,\Omega}{2\pi \nu \Sigma r}
+
\frac{\dot J}{2\pi \nu \Sigma r^3},
\end{equation}
where \(\Omega\) is the angular velocity, \(\dot m\) is the mass inflow rate, \(\nu\) is the kinematic viscosity, and \(\Sigma\) is the surface density.

For a purely Keplerian disc one has
\begin{equation}
\Omega_{\mathrm{Kep}}=\sqrt{\frac{GM}{r^3}},
\end{equation}
and the standard solution gives
\begin{equation}
\nu\Sigma
=
\frac{\dot m}{3\pi}
\left(
1-\frac{\dot J}{\dot m\sqrt{GMr}}
\right).
\end{equation}
However, that treatment does not explicitly incorporate the fact that the star itself may be rotating close to the Keplerian value at the inner edge of the disc. This motivates the search for a near-critical solution of the type found by \citetalias{Paczynski1991} and \citetalias{Popham1991}, in which the disc connects continuously to the stellar surface.

\subsection{Basic approximation}

We are interested specifically in the regime where the stellar surface is near critical rotation, so that
\begin{equation}
\Omega^2-\Omega_{\mathrm{Kep}}^2
\end{equation}
remains small throughout the inner disc. In that regime the radial momentum equation simplifies considerably. Starting from
\begin{equation}
u_r \frac{\partial u_r}{\partial r}
=
-\frac{1}{\rho}\frac{\partial P}{\partial r}
+r\left(\Omega^2-\Omega_{\mathrm{Kep}}^2\right),
\end{equation}
where $u_r < 0$ for accretion, and we neglect the final term to leading order and solve the remaining problem perturbatively. The purpose of the derivation is therefore not to describe arbitrary discs, but rather to isolate an analytic near-critical solution relevant to our application.

The approach is as follows. First, we derive the vertical structure of a polytropic disc and obtain the radial scalings of \(\rho\), \(\Sigma\), \(c_s\), and \(\nu\). We then show that there exists a self-consistent solution for which \(\nu\Sigma\) is independent of \(r\). Once that has been established, the angular-momentum equation can be integrated directly.

\subsection{Vertical structure of a polytropic disc}

Assume a polytropic equation of state,
\begin{equation}
P=K\rho^{1+1/n},
\end{equation}
where $K$ is a constant and \(n\) is the polytropic index. Vertical hydrostatic balance gives
\begin{equation}
\frac{1}{\rho}\frac{\partial P}{\partial z}
=
-\frac{GMz}{r^3}.
\end{equation}
Substituting the polytropic equation of state,
\begin{equation}
\frac{1}{\rho}\frac{\partial \rho}{\partial z}
K\frac{n+1}{n}\rho^{1/n}
=
-\frac{GMz}{r^3},
\end{equation}
or equivalently
\begin{equation}
\partial_z\!\left[K(n+1)\rho^{1/n}\right]
=
-\frac{GMz}{r^3}.
\end{equation}

Integrating with respect to \(z\), and defining \(z_h(r)\) as the disc half-thickness at cylindrical radius \(r\), 
we obtain
\begin{equation}
K(n+1)\rho^{1/n}
=
-\frac{GM}{r^3}\left[\frac{z^2}{2}\right]_{z_h}^{z}
=
-\frac{GM}{2r^3}\left(z^2-z_h^2\right).
\end{equation}
Hence
\begin{equation}
\rho^{1/n}
=
\frac{GM}{2K(n+1)r^3}\left(z_h^2-z^2\right),
\end{equation}
and therefore
\begin{equation}
\rho
=
\left[\frac{GM}{2K(n+1)r^3}\right]^n
\left(z_h^2-z^2\right)^n.
\label{apeq:rho}
\end{equation}

\subsection{Surface density}

The surface density is
\begin{equation}
\Sigma=\int_{-z_h}^{z_h}\rho\,dz.
\end{equation}
Substituting Eq.\,\ref{apeq:rho} into the expression above,
\begin{equation}
\Sigma
=
\left[\frac{GM}{2K(n+1)r^3}\right]^n
\int_{-z_h}^{z_h}\left(z_h^2-z^2\right)^n\,dz.
\end{equation}
The remaining integral is standard and can be expressed in terms of Gamma functions, yielding:
\begin{equation}
\int_{-z_h}^{z_h}\left(z_h^2-z^2\right)^n\,dz
=
\sqrt{\pi}\,z_h^{2n+1}\,
\frac{\Gamma(n+1)}{\Gamma(n+3/2)}.
\end{equation}
Thus
\begin{equation}
\Sigma
=
\left[\frac{GM}{2K(n+1)}\right]^n
\frac{\sqrt{\pi}\,\Gamma(n+1)}{\Gamma(n+3/2)}
\frac{z_h^{2n+1}}{r^{3n}}.
\label{apeq:Sigma}
\end{equation}
For the radial scaling we therefore have
\begin{equation}
\Sigma \propto \frac{z_h^{2n+1}}{r^{3n}}.
\end{equation}

\subsection{Sound speed and viscosity}

Using again vertical hydrostatic balance,
\begin{equation}
\frac{1}{\rho}\frac{\partial P}{\partial z}
=
\frac{\partial \ln\rho}{\partial z}\,c_s^2
=
-\frac{GMz}{r^3},
\end{equation}
where the sound speed \(c_s^2=\partial P/\partial\rho\). Since
\begin{equation}
\rho \propto (z_h^2-z^2)^n,
\end{equation}
we have
\begin{equation}
\frac{\partial \ln\rho}{\partial z}
=
n\,\frac{\partial}{\partial z}\ln(z_h^2-z^2)
=
-\frac{2nz}{z_h^2-z^2}.
\end{equation}
Substituting this into the hydrostatic equation gives
\begin{equation}
c_s^2
=
\frac{GM}{2r^3}\left(z_h^2-z^2\right).
\end{equation}
At the mid-plane \(z=0\),
\begin{equation}
c_s^2(z=0)=\frac{GMz_h^2}{2r^3},
\end{equation}
so that
\begin{equation}
c_s \propto \frac{z_h}{r^{3/2}}.
\end{equation}

Using the \(\alpha\)-disc model for viscosity \citep{Shakura1973},
\begin{equation}
\nu=\alpha c_s z_h,
\end{equation}
where $\alpha$ is a dimensionless parameter, we obtain
\begin{equation}
\nu \propto \frac{z_h^2}{r^{3/2}}.
\end{equation}

Combining this with the expression for \(\Sigma\) from Eq.\,\ref{apeq:Sigma},
\begin{equation}
\nu\Sigma \propto \frac{z_h^{2n+3}}{r^{3n+3/2}}.
\label{apeq:nuSigma}
\end{equation}

\subsection{Condition for \texorpdfstring{\(\nu\Sigma\)}{nu Sigma} to be constant}

We now ask whether there exists a self-consistent power-law solution with
\begin{equation}
z_h \propto r^\beta
\end{equation}
such that \(\nu\Sigma\) is independent of \(r\). Substituting this into Eq.\,\ref{apeq:nuSigma} gives
\begin{equation}
\nu\Sigma \propto r^{\beta(2n+3)-(3n+3/2)}.
\end{equation}
Requiring \(\nu\Sigma\propto r^0\) yields
\begin{equation}
\beta=\frac{3n+3/2}{2n+3}.
\label{apeq:betafirst}
\end{equation}

At this stage this is only one condition. We now use the radial momentum equation to obtain a second relation between \(n\) and \(\beta\).

\subsection{Constraint from the radial momentum equation}

Starting from
\begin{equation}
\rho u_r \frac{\partial u_r}{\partial r}
=
-\frac{\partial}{\partial r}\!\left(K\rho^{1+1/n}\right)
-\rho r\Omega_{\mathrm{Kep}}^2
+\rho r\Omega^2,
\end{equation}
we use mass conservation,
\begin{equation}
\Sigma u_r=-\frac{\dot m}{2\pi r},
\end{equation}
to write
\begin{equation}
\rho\frac{\dot m^2}{4\pi^2 r\Sigma}
\frac{\partial}{\partial r}\!\left(\frac{1}{r\Sigma}\right)
=
-K\frac{\partial}{\partial r}\!\left(\rho^{1+1/n}\right)
-\rho r\Omega_{\mathrm{Kep}}^2
+\rho r\Omega^2.
\end{equation}
Integrating over \(z\), and using \(\int \rho\,dz=\Sigma\), gives
\begin{equation}
\frac{\dot m^2}{4\pi^2 r}
\frac{\partial}{\partial r}\!\left(\frac{1}{r\Sigma}\right)
=
-\int_{-z_h}^{z_h}
K\frac{\partial}{\partial r}\!\left(\rho^{1+1/n}\right)\,dz
-\Sigma r\left(\Omega^2-\Omega_{\mathrm{Kep}}^2\right).
\label{apeq:radmom}
\end{equation}

To extract only the radial scaling, substitute
\begin{equation}
\rho \propto \frac{(z_h^2-z^2)^n}{r^{3n}},
\qquad
z_h\propto r^\beta.
\end{equation}
Then the integral term scales as
\begin{equation}
\int_{-z_h}^{z_h}
\frac{\partial}{\partial r}
\left[
\frac{(z_h^2-z^2)^{n+1}}{r^{3n+3}}
\right]dz
\propto
\frac{z_h^{2n+3}}{r^{3n+4}}
\propto
\frac{r^{2n\beta+3\beta}}{r^{3n+4}}.
\end{equation}

On the left-hand side of Eq.\,\ref{apeq:radmom} we use
\begin{equation}
\Sigma \propto \frac{z_h^{2n+1}}{r^{3n}}
\propto
\frac{r^{2n\beta+\beta}}{r^{3n}},
\end{equation}
so that
\begin{equation}
\frac{1}{r}\frac{\partial}{\partial r}\left(\frac{1}{r\Sigma}\right)
\propto
\frac{r^{-2n\beta-\beta}}{r^{-3n+3}}.
\end{equation}

Now we invoke the key near-critical assumption of this appendix: the term proportional to \(\Omega^2-\Omega_{\mathrm{Kep}}^2\) is small compared to the other terms. Then the left- and right-hand sides of Eq.\,\ref{apeq:radmom} must have the same radial scaling, which implies
\begin{equation}
-2n\beta-\beta+3n-3 = 2n\beta+3\beta-3n-4.
\end{equation}
This simplifies to
\begin{equation}
6n+1 = 4\beta n + 4\beta,
\end{equation}
or
\begin{equation}
\beta=\frac{6n+1}{4n+4}.
\end{equation}

We now equate this with the previous expression for \(\beta\) from Eq.\,\ref{apeq:betafirst}:
\begin{equation}
\frac{3n+3/2}{2n+3}=\frac{6n+1}{4n+4}.
\end{equation}
Solving gives
\begin{equation}
n=\frac{3}{2}.
\end{equation}
Substituting back,
\begin{equation}
\beta=1.
\end{equation}

We therefore obtain the scalings
\begin{align}
z_h &\propto r,\\
c_s &\propto r^{-1/2},\\
\Sigma &\propto r^{-1/2},\\
\nu &\propto r^{1/2},
\end{align}
and hence
\begin{equation}
\nu\Sigma \propto r^0.
\end{equation}

This recovers the disc-height scaling \(z_h\propto r\) assumed by \citetalias{Paczynski1991}, equivalent to constant \(H/R\), and also gives \(c_s\propto r^{-1/2}\), in agreement with the scaling used by \citetalias{Martin2025}.

\subsection{Range of validity of the approximation}

The derivation above relies on neglecting the term proportional to \(\Omega^2-\Omega_{\mathrm{Kep}}^2\) in the radial momentum equation. To test whether this is at least plausible in the regime relevant to our application, we evaluate the terms for one representative accretor in our simulations, with
\begin{equation}
M_2 = 7.64\,M_\odot,\qquad
R_2 = 5.44\,R_\odot,\qquad
\dot m = 5.3\times 10^{-5}\,M_\odot\,{\rm yr}^{-1},
\end{equation}
and adopt \(H/R=0.2\), similar to one of the thicker discs considered by \citetalias{Martin2025}. Using the density and radial gradients at the stellar surface, together with the disc sound speed, we obtain
\begin{equation}
4.88\times 10^{-7}
-
2.62\times 10^{-7}
=
1.91\times 10^{-8}
\left(1-\frac{\Omega^2}{\Omega_{\mathrm{Kep}}^2}\right).
\end{equation}
For large deviations from Keplerian rotation the right-hand side is not negligible. However, if
\begin{equation}
\left|1-\frac{\Omega^2}{\Omega_{\mathrm{Kep}}^2}\right|\approx 0.13,
\end{equation}
then the rotational term contributes less than \(1\,\%\) of the magnitude of the two leading terms in the radial momentum equation. This corresponds to approximately
\begin{equation}
0.93 \le \frac{\Omega}{\Omega_{\mathrm{Kep}}} \le 1.066.
\end{equation}
This is a narrow but relevant regime for our intended application, namely stars that are at or only slightly above critical rotation. A broader exploration of the validity domain is left to future work.

\subsection{Solution of the angular-momentum equation}

Having established that a self-consistent solution exists with \(\nu\Sigma\approx\) const, we now return to the angular-momentum equation:
\begin{equation}
\frac{\partial \Omega}{\partial r}
+
\frac{\dot m}{2\pi \nu\Sigma r}\Omega
=
\frac{\dot J}{2\pi \nu\Sigma r^3}.
\end{equation}
Since \(\nu\Sigma\) is constant, this is a linear first-order ordinary differential equation in \(\Omega(r)\).
We multiply by an integrating factor $r^{\dot m/(2\pi \nu\Sigma)}$,
\begin{equation}
\frac{\partial}{\partial r}
\left[
\Omega\,r^{\dot m/(2\pi \nu\Sigma)}
\right]
=
\frac{\dot J}{2\pi \nu\Sigma}\,
r^{\dot m/(2\pi \nu\Sigma)-3}.
\end{equation}
Integrating,
\begin{equation}
\Omega\,r^{\dot m/(2\pi \nu\Sigma)}
=
\frac{\dot J}{2\pi \nu\Sigma\left(\dot m/(2\pi \nu\Sigma)-2\right)}
\,r^{\dot m/(2\pi \nu\Sigma)-2}
+
C.
\end{equation}
Therefore
\begin{equation}
\Omega
=
\frac{\dot J}{2\pi \nu\Sigma\left(\dot m/(2\pi \nu\Sigma)-2\right)}
\,r^{-2}
+
C\,r^{-\dot m/(2\pi \nu\Sigma)}.
\end{equation}

To recover a Keplerian outer disc at large radii, the second term must scale as \(r^{-3/2}\). This requires
\begin{equation}
\frac{\dot m}{2\pi \nu\Sigma}=\frac{3}{2},
\end{equation}
which is equivalent to
\begin{equation}
\nu\Sigma=\frac{\dot m}{3\pi}.
\end{equation}
The solution then becomes
\begin{equation}
\Omega
=
-\frac{3\dot J}{\dot m r^2}
+
C\,r^{-3/2}.
\end{equation}
Requiring the outer disc to approach \(\Omega_{\mathrm{Kep}}=\sqrt{GM/r^3}\) fixes
\begin{equation}
C=\sqrt{GM},
\end{equation}
and hence
\begin{equation}
\Omega(r)=\Omega_{\mathrm{Kep}}-\frac{3\dot J}{\dot m r^2}.
\end{equation}

Thus the difference between the Keplerian angular velocity and the actual disc angular velocity is
\begin{equation}
\Omega_{\mathrm{Kep}}-\Omega
=
\frac{3\dot J}{\dot m r^2}.
\end{equation}
Evaluated at the star--disc boundary,
\begin{equation}
\dot J
=
\left(\Omega_{\mathrm{Kep}}(R_\star)-\Omega_\star\right)
\frac{\dot m R_\star^2}{3}.
\end{equation}

This is the key result presented in Section\,\ref{sec:ADM}. As \(\Omega_\star\) approaches \(\Omega_{\mathrm{Kep}}(R_\star)\), the torque tends to zero. If the stellar surface is perturbed to super-critical rotation, \(\Omega_\star>\Omega_{\mathrm{Kep}}(R_\star)\), then \(\dot J <0\): the disc extracts angular momentum from the star while \(\dot m\) can remain positive. This is precisely the behaviour required for continued accretion near critical rotation.

\subsection{Equivalent form in terms of specific angular momentum}

It is sometimes useful to write the torque per unit mass flux as
\begin{equation}
\dot J = \frac{\dot J}{\dot m}.
\end{equation}
In the present model,
\begin{equation}
\dot J
=
\left(\Omega_{\mathrm{Kep}}(R_\star)-\Omega_\star\right)\frac{R_\star^2}{3}.
\end{equation}
For sub-critical rotation this is positive, so accretion spins the star up. For super-critical rotation the right-hand side becomes negative, and the disc removes angular momentum from the star even though the net mass flux remains inward.

\subsection{Recovering the condition \texorpdfstring{\(\nu\Sigma=\dot m/(3\pi)\)}{nu Sigma = mdot/(3pi)}}

The condition \(\dot m/(2\pi \nu\Sigma)=3/2\) was introduced above by requiring the outer solution to be Keplerian. One can also recover it directly by substituting the solution for \(\Omega\) back into the angular-momentum equation.

Take
\begin{equation}
\Omega=\Omega_{\mathrm{Kep}}-\frac{3\dot J}{\dot m r^2}.
\end{equation}
Then
\begin{equation}
\nu\Sigma\frac{\partial \Omega}{\partial r}
=
-\frac{\dot m\Omega}{2\pi r}
+
\frac{\dot J}{2\pi r^3}.
\end{equation}
Substituting for \(\Omega\) and simplifying yields
\begin{equation}
\nu\Sigma
\left(
-\frac{\sqrt{GM}}{2r^{5/2}}
-\frac{2\dot J}{\dot m r^3}
\right)
=
\frac{\dot m}{3\pi}
\left(
\frac{\sqrt{GM}}{2r^{5/2}}
+\frac{2\dot J}{\dot m r^3}
\right),
\end{equation}
from which it follows that
\begin{equation}
\nu\Sigma=\frac{\dot m}{3\pi}.
\end{equation}
This is somewhat circular, since the same solution has already been assumed, but it shows explicitly that the Keplerian outer-boundary condition is equivalent to the usual \(\nu\Sigma\) relation.

\subsection{Comparison with previous work}

A useful point of comparison is that \citetalias{Paczynski1991} and \citetalias{Popham1991} adopt polytropic disc models, whereas \citetalias{Martin2025} instead assumes constant \(H/R\). In our derivation these are not in conflict: once one obtains \(n=3/2\), the result \(z_h\propto r\) follows immediately, which is equivalent to constant \(H/R\). Likewise,
\begin{equation}
c_s \propto \frac{H}{r}\,r\Omega_{\mathrm{Kep}}
\propto r^{-1/2},
\end{equation}
in agreement with both our result and the scaling used by \citetalias{Martin2025}.

Although our scale height \(z_h(r)\) represents a sharp boundary in this polytropic treatment, it can be mapped to a more physically realistic Gaussian profile $\rho\,_\mathrm{Gaussian}\propto \mathrm{exp}(-\frac{z^2}{2\,z_{h}^2})$ used in various disc models, for which the polytropic scaling height maps to the standard deviation as $z_h\propto\sigma\propto\frac{c_s}{\Omega}$.

The present derivation therefore connects the near-critical polytropic solutions of \citetalias{Paczynski1991} to the constant-\(H/R\) picture used for Be-star decretion discs. In our case, however, we are interested in an accretion flow with \(\dot m>0\). The important point is that a negative \(\dot J\) can still arise even when the mass flux remains inward, provided the stellar surface rotates slightly faster than the Keplerian value at the inner edge of the disc.

\section{Inlist settings}
\label{section:appendixInlist}

We list here a representative subset of the inlist parameters used in our calculations. These entries are intended to illustrate the main numerical choices and do not constitute complete inlists. The same inlist was used for both primary and secondary stars. Full inlists will be made available with the published data.

\subsection{Star settings}

\begin{verbatim}
Zbase = 0.0185
v_flag = .true.
dump_missing_metals_into_heaviest = .false.
reaction_net = 'approx21.net'
use_Type2_opacities = .true.
kap_file_prefix = 'a09'
kap_lowT_prefix = 'lowT_fa05_a09p'

atm_option = 'T_tau'
Pextra_factor = 0

mixing_length_alpha = 2d0
MLT_option = 'Henyey'
do_conv_premix = .true.
use_Ledoux_criterion = .true.
use_superad_reduction = .true.
alpha_semiconvection = 100
semiconvection_option = 'Langer_85 mixing; gradT = gradr'
thermohaline_coeff = 1
thermohaline_option = 'Kippenhahn'

varcontrol_target = 7d-4
mesh_delta_coeff = 0.8
max_dq = 3d-3
T_function2_weight = 180
T_function2_param = 1d5
min_dq_for_xa = 1d-6
premix_omega = .true.
recalc_mixing_info_each_substep = .true.
mesh_min_dlnR = 3d-11
restore_mesh_on_retry = .true. 
mdot_omega_power = 0
limit_for_rel_error_in_energy_conservation = 3d-6
dt_div_min_dr_div_cs_limit = 1d-2

adjust_J_fraction = 0.4d0
include_rotation_in_total_energy = .true.
fp_rot = 0.8   (implemented via x_ctrl(10))
\end{verbatim}

\subsection{Binary settings}

\begin{verbatim}
mdot_scheme = 'Kolb'

do_wind_mass_transfer = .false.
do_enhance_wind = .true.
do_jdot_mb = .false.
do_tidal_sync = .true.
do_initial_orbit_sync = .true.
do_tidal_circ = .true.

CE_begin_at_max_implicit_abs_mdot = .false.
\end{verbatim}

\section{SMC metallicity distributions}
\label{section:appendixSMC}

\begin{figure*}
    \centering    
    \begin{minipage}{0.48\textwidth}
    \hspace{-5mm}
    \includegraphics[width=1.08\linewidth,trim={0 0 0 0.0cm},clip]{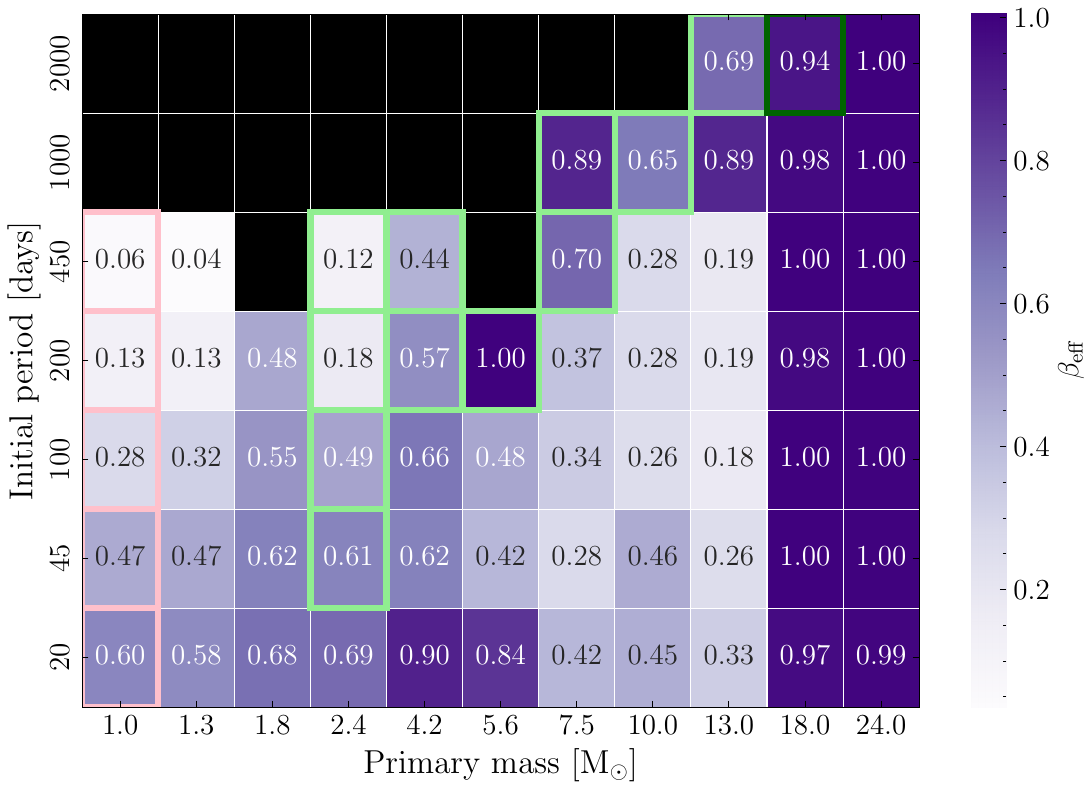}
    \end{minipage}
    \hfill
    \begin{minipage}{0.48\textwidth}
    \hspace{-2mm}
    \includegraphics[width=1.08\linewidth,trim={0 0 0 0.0cm},clip]{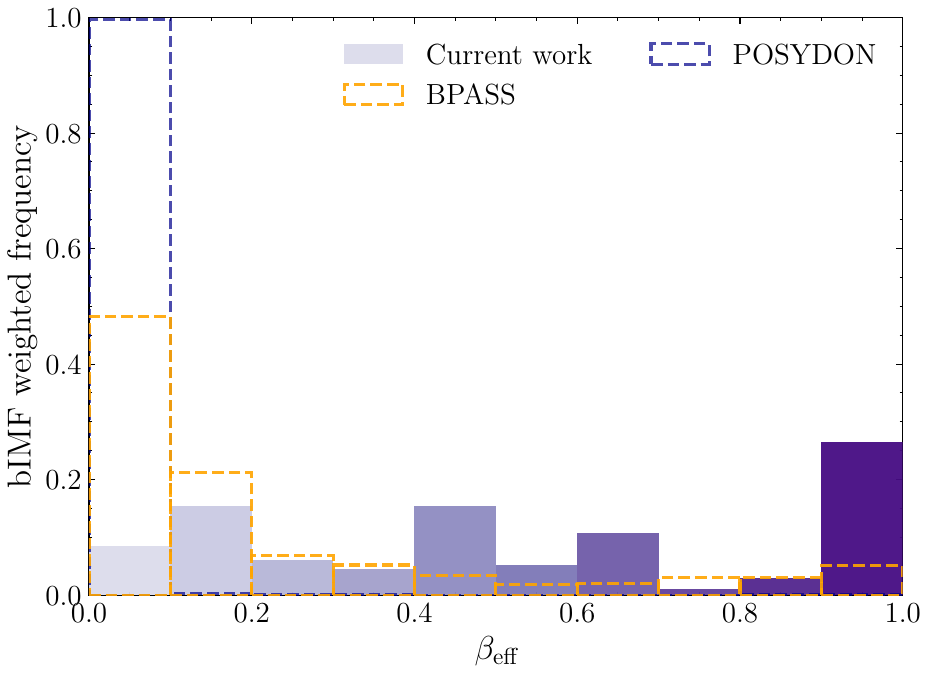}\vspace{-0.0cm}
    \end{minipage}
    \caption{\textit{Left:} A heatmap of $\beta_\mathrm{eff}$ for a binary mass ratio of $0.8$ at SMC metallicity, across initial periods and primary masses. The black squares represent cases where negligible mass-transfer occurs ($\dot{m}_\mathrm{negligible} < 10^{-10}\, \mathrm{M}_{\odot}\,\mathrm{yr}^{-1}$). We highlight models that undergo case C (light green) mass transfer, or where the accretor has a convective envelope (light pink) at the start of mass transfer. \textit{Right:} Binary-fraction and IMF-weighted histogram for binaries, for the same q= 0.8 parameter space as shown in the heatmap at SMC metallicity, comparing the disc model simulations with \textsc{bpass} and \textsc{posydon} grids. The resulting distributions are similar to their Solar metallicity counterparts, which for the current work shows a flatter distribution than that of \textsc{bpass} or \textsc{posydon}.}
    \label{fig:beff_grid_SMC}
\end{figure*}
Here we present additional results at a lower metallicity appropriate for the SMC, $Z=\frac{\mathrm{Z}_{\odot}}{4}$. At SMC metallicity, the overall behaviour remains qualitatively similar to that found for the Solar-metallicity grid. Fig.\,\ref{fig:beff_grid_SMC} shows that the disc model produces a broad range of effective mass-transfer efficiencies, with the IMF-weighted distribution being flat whilst a bottom heavy distribution is seen for \textsc{bpass}, and \textsc{posydon} remains strongly peaked at $\beta_\mathrm{eff}<10\,\%$. In particular, the disc models retain a substantial contribution from systems with intermediate-to-high mass-transfer efficiencies, consistent with the range favoured by the SMC Be-star population discussed in Section\,\ref{subsec:IFM}.

\begin{figure*}
    \centering    
    \begin{minipage}{0.48\textwidth}
    \hspace{-5mm}
    \includegraphics[width=1.05\linewidth,trim={0 0 0 0.0cm},clip]{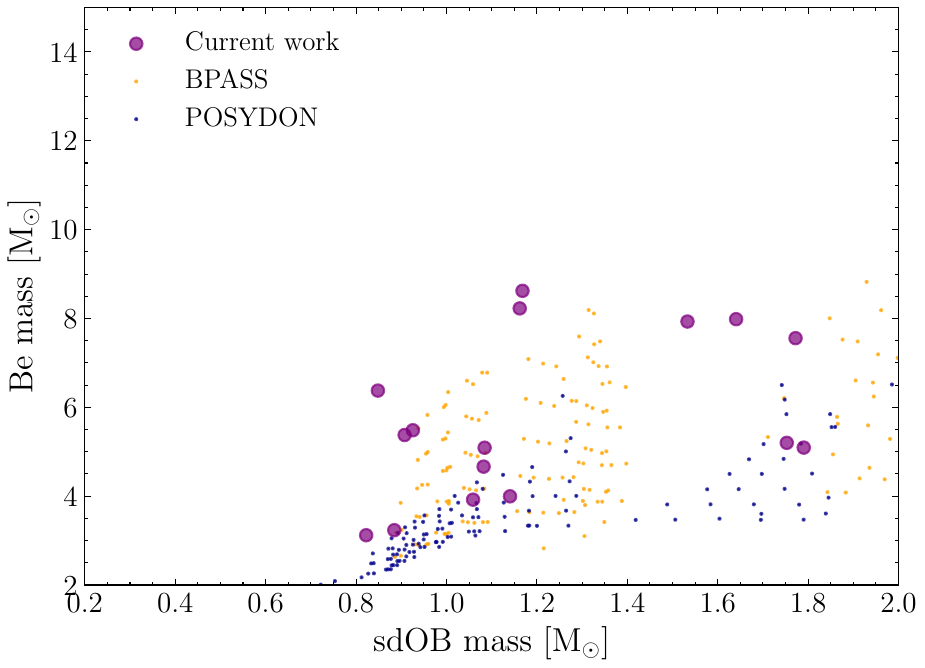}
    \end{minipage}
    \hfill
    \begin{minipage}{0.48\textwidth}
    \hspace{-2mm}
    \includegraphics[width=1.05\linewidth,trim={0 0 0 0.0cm},clip]{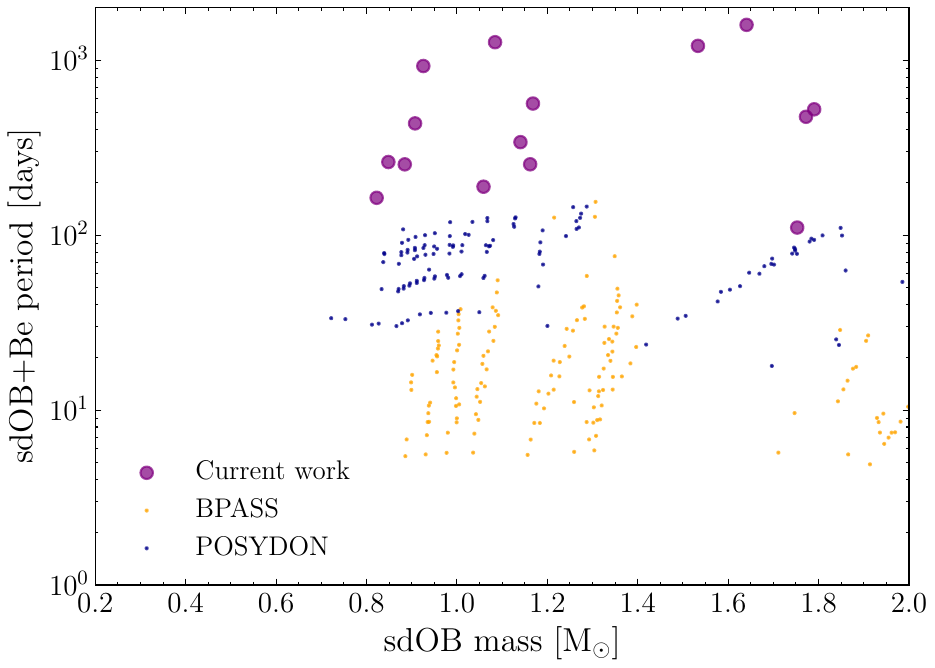}\vspace{-0.0cm}
    \end{minipage}
    \caption{The post-MT properties of sdOB ($10000< T_{\mathrm{eff}, 1}$, $\mathrm{surface\_hydrogen}<0.5$) + Be ($10000< T_{\mathrm{eff}, 2}<40000\,\mathrm{K}$, $0.7< {\Omega_\mathrm{surface}} / {\Omega_{\mathrm{crit}}}$) binaries for \textsc{bpass} (excluding the rotation filter), \textsc{posydon}, and the disc model at SMC metallicity. We note that the SMC metallicity grid only includes binaries that underwent case B or case C mass transfer. \textit{Left:} The post-MT Be against sdOB masses. \textit{Right:} The post-MT periods against sdOB masses at SMC metallicity.}
    \label{fig:BeHe_SMC}
\end{figure*}
Fig.\,\ref{fig:BeHe_SMC} shows the corresponding post-mass-transfer sdOB+Be population, allowing the component-mass and orbital-period distributions to be compared directly with the Solar-metallicity results discussed in Section\,\ref{subsec:CLE}. In the Solar metallicity grid, we found that not including case A contributions leads to the absence of lower sdOB masses below $0.8\,\mathrm{M}_\odot$. The absence of these lower sdOB masses in this SMC sample likely corresponds to the lack of case A regime periods in the SMC metallicity grid. Overall, in the compared parameter space, the disc model at SMC metallicity results in slightly higher relative Be masses and larger separations.

%%%%%%%%%%%%%%%%%%%%%%%%%%%%%%%%%%%%%%%%%%%%%%%%%%

\bsp	% typesetting comment
\label{lastpage}
\end{document}